\documentclass[5p,twocolumn,authoryear,times]{elsarticle}
\usepackage{lineno,hyperref}
\modulolinenumbers[5]
\hypersetup{colorlinks,citecolor=blue,linkcolor=blue,urlcolor=blue}

\usepackage{graphicx}	
\usepackage{amsmath}	
\usepackage{amssymb}	
\usepackage{bm}

\usepackage{natbib} 
\usepackage[usenames,dvipsnames]{color}
\usepackage{verbatim}
\usepackage{multirow}
\usepackage{booktabs}
\usepackage{subfigure}
\usepackage{color, colortbl}

\usepackage[T1]{fontenc}
\usepackage[utf8]{inputenc}
\usepackage[tracking=smallcaps]{microtype}
\SetTracking[no ligatures = f]{encoding = *,shape = sc}{0}

\usepackage{listings}
\usepackage[normalem]{ulem}
\usepackage[font=footnotesize,labelfont=bf]{caption}
\usepackage{footnote}
\makesavenoteenv{tabular}
\makesavenoteenv{table}
\usepackage{supertabular}
\usepackage{threeparttable}
\usepackage{longtable}
\usepackage{rotating}
\usepackage{lscape}

\usepackage{lineno}

\usepackage{bold-extra}
\usepackage{slantsc}
\usepackage{times}

\usepackage{blindtext}
\makeatletter
\newcommand{\removelatexerror}{\let\@latex@error\@gobble}
\makeatother

\usepackage[]{algorithm2e}
\usepackage{siunitx}
\usepackage{dsfont}
\usepackage{bbm}

\usepackage{tikz,pgfplots}

\let\oldFootnote\footnote
\newcommand\nextToken\relax

\renewcommand\footnote[1]{%
    \oldFootnote{#1}\futurelet\nextToken\isFootnote}

\newcommand\isFootnote{%
    \ifx\footnote\nextToken\textsuperscript{,}\fi}
\usepackage[multiple]{footmisc}

\definecolor{babypink}{rgb}{0.96, 0.76, 0.76}
\definecolor{beaublue}{rgb}{0.74, 0.83, 0.9}
\definecolor{mediumturquoise}{rgb}{0.28, 0.82, 0.8}
\definecolor{mossgreen}{rgb}{0.68, 0.87, 0.68}
\definecolor{mustard}{rgb}{1.0, 0.86, 0.35}
\definecolor{olivine}{rgb}{0.6, 0.73, 0.45}
\definecolor{orchid}{rgb}{0.85, 0.44, 0.84}
\definecolor{palecerulean}{rgb}{0.61, 0.77, 0.89}
\definecolor{palegold}{rgb}{0.9, 0.75, 0.54}
\definecolor{paleplum}{rgb}{0.8, 0.6, 0.8}
\definecolor{palespringbud}{rgb}{0.93, 0.92, 0.74}
\definecolor{pastelgray}{rgb}{0.81, 0.81, 0.77}
\definecolor{pastelviolet}{rgb}{0.8, 0.6, 0.79}
\definecolor{pastelred}{rgb}{1.0, 0.41, 0.38}
\definecolor{pearl}{rgb}{0.94, 0.92, 0.84}
\definecolor{pistachio}{rgb}{0.58, 0.77, 0.45}	
\definecolor{teal}{rgb}{0.0, 0.5, 0.5}
\definecolor{tiffanyblue}{rgb}{0.04, 0.73, 0.71}
\definecolor{turquoise}{rgb}{0.19, 0.84, 0.78}
\definecolor{verdigris}{rgb}{0.26, 0.7, 0.68}
\definecolor{lightgray}{gray}{0.9}

\definecolor{blue_mlp}{rgb}{0.122, 0.467, 0.706}
\definecolor{orange_mlp}{rgb}{0.957, 0.643, 0.376}
\definecolor{red_mlp}{rgb}{0.804, 0.361, 0.361}
\definecolor{green_mlp}{rgb}{0.09, 0.723, 0.068}

\newcommand{\maasai}{\textsc{maasai}}

 {\pagebreak[4]\global\pdfpageattr\expandafter{\the\pdfpageattr/Rotate 90}}%
 {\pagebreak[4]\global\pdfpageattr\expandafter{\the\pdfpageattr/Rotate 0}}%
 
\usepackage{listings}
\lstnewenvironment{codebkg}{%
  \lstset{
    escapechar=\&,
    basicstyle=\ttfamily\scriptsize,
    backgroundcolor=\color{lightgray},
    frame=single,
    breaklines=true,    
    framerule=0pt,
    columns=fullflexible
  }
}
{}

\lstnewenvironment{texthighlight}{%
  \lstset{
    escapechar=\&,
    basicstyle=\ttfamily\scriptsize,
    backgroundcolor=\color{lightgray},
    frame=single,
    breaklines=true,  
    breakindent=0em,
    framerule=0pt,
    columns=fullflexible
  }
  \lstset{
    moredelim=[is][\itshape]{**}{**}
  }
}
{}

\begin{document}
\begin{frontmatter}

\title{MAASAI: A Tool-Augmented Multi-Agent System for Astronomical Data Analysis}

\author[1]{S.~Riggi\corref{cor}}%
\ead{simone.riggi@inaf.it}
\author[1]{A.~Ingallinera}
\author[2]{C.~Bordiu}
\author[1]{C.~Buemi}
\author[1]{F.~Bufano}
\author[3]{M.~Burgay}
\author[1]{F.~Cavallaro}
\author[1]{\\P.~Leto}
\author[1]{S.~Loru}
\author[4]{S.E.~Motta}
\author[3]{M.~Pilia}
\author[4]{T.~Sbarrato}

\cortext[cor]{Corresponding author}%
\address[1]{INAF - Osservatorio Astrofisico di Catania, Via Santa Sofia 78, 95123 Catania, Italy}%
\address[2]{Instituto de Astrofísica de Andalucía (CSIC), Glorieta de la Astronomía s/n, 18008 Granada, Spain}%
\address[3]{INAF - Osservatorio Astronomico di Cagliari, Via della Scienza 5, 09047 Selargius (CA), Italy}%
\address[4]{INAF - Osservatorio Astronomico di Brera, Via Bianchi 46, 23807 Merate (LC), Italy}%

\begin{abstract}
Modern astronomical workflows increasingly require the coordinated use of specialised software, scientific archives, heterogeneous data products, and domain-specific knowledge. Recent advances in large language models and multimodal foundation models offer new opportunities for natural-language interaction with scientific data and computational tools. Realising this potential in astronomy, however, requires integrating these models with established scientific capabilities.

We present \maasai{} (Multi-agent Astronomy Assistant for Scientific Analysis and Inference), a modular multi-agent system combining foundation-model reasoning, retrieval-augmented scientific knowledge, and specialised astronomical tools within a graph-based execution framework. \maasai{} coordinates request preparation, task planning, execution, and response synthesis through specialised workers. Its architecture supports native and external scientific applications, structured artefact and provenance management, optional human approval, and configurable inference backends, including fully local deployment with open-weight models. Motivated primarily by the data-analysis challenges of the SKA and its precursor facilities, \maasai{} is also designed to support applications in other astronomical domains.

We evaluate \maasai{} through radio-image benchmarks, single- and multi-step workflow demonstrations, and literature-grounded reasoning. Compared with a general-purpose commercial-model baseline, \maasai{} improves macro-averaged F1 scores from 0.53 to 0.72 for extended/diffuse emission classification and from 0.37 to 0.88 for source morphology classification. Workflow demonstrations illustrate the coordination of astronomical measurements, source extraction, archive queries, and catalogue processing. A 94-question literature benchmark yielded generally favourable automated assessments, with comparable ratings from a pilot expert evaluation of 59 answers. These results highlight how tool-augmented multi-agent systems can extend foundation-model reasoning to specialised astronomical analyses by integrating established scientific software within coordinated workflows.
\end{abstract}


\end{frontmatter}

\section{Introduction}
\label{sec:intro}
Astronomy is undergoing a profound transformation driven by the unprecedented volume and complexity of data produced by modern observational facilities. In radio astronomy, surveys carried out by the \textit{Australian SKA Pathfinder} (ASKAP) \citep{Johnston2008,Hotan2021}, \textit{MeerKAT} \citep{Jonas2016}, and the \textit{LOw-Frequency ARray} (LOFAR) \citep{vanHaarlem2013} telescopes are producing catalogues containing millions of sources and petabytes of imaging data, with further growth expected from the forthcoming \textit{Square Kilometre Array} (SKA) \citep{SKADesignDoc}. , with further growth expected from the forthcoming \textit{Square Kilometre Array} (SKA) \citep{SKADesignDoc}. Similar trends are occurring across other astronomical domains, involving both ground-based facilities in the optical and gamma-ray bands, such as the \textit{Vera C. Rubin Observatory} \citep{Ivezić2019} and the \textit{Cherenkov Telescope Array Observatory} (CTAO) \citep{CTA,Hofmann2024}, and space missions \citep{Mellier2025,SDO,Muller2020,Ricker2015,Rauer2025}. The resulting scientific products---including images, spectral cubes, catalogues, spectra, and light curves---increasingly require automated methods capable of extracting scientific information efficiently and at scale.

Over the past decade, deep learning has become a key technology for addressing the challenges of data-intensive astronomy across a wide range of scientific applications \citep{YuanSenTing2026,Sahakyan2025}. These advances have produced a growing ecosystem of specialised models and software tools, often developed for specific tasks, surveys, or instruments. However, their use frequently requires substantial technical expertise, and limited integration into broader scientific workflows can hinder their accessibility to the astronomical community.

More recently, large language models (LLMs) and multimodal foundation models have opened new possibilities for scientific analysis through natural-language interaction. In astronomy, applications now span scientific question answering and knowledge assistance \citep{Nguyen2023,Perkowski2024,AstroMLab3,AstroMLab4}, image understanding and multimodal retrieval \citep{PAPERCLIP,AstroCLIP,Riggi2025,Gupta2025,Stoppa2025}, and literature-grounded reasoning through retrieval-augmented generation \citep{Iyer2024,YuanSenTing2025,Xu2025,Wang2026,Asai2026}. 
These developments are now extending towards agentic AI systems, which combine foundation-model reasoning with task planning, knowledge retrieval, and external tool execution to support multi-step scientific workflows (Section~\ref{sec:literature-review}).

Despite these advances, the effectiveness of general-purpose multimodal foundation models for specialised astronomical data analysis remains highly dependent on the task and prompting strategy \citep{Riggi2025,Stoppa2025,Drozdova2025}. Domain adaptation of small models through instruction tuning can improve performance, but is constrained by the scarcity and limited diversity of annotated astronomical datasets, the risk of catastrophic forgetting, and the fact that adapted models may still underperform pure vision-only models \citep{Riggi2025}.

These limitations motivate an alternative approach in which agentic systems combine general-purpose foundation models with established astronomical tools and knowledge resources, rather than embedding every specialised scientific capability into the models themselves. Such an approach enables the integration of community-developed software, institution-specific services, and specialised applications that are not readily accessible through general-purpose AI platforms. Scientific computations can thus be delegated to tools and machine-learning models with independently evaluated performance and, where available, quantified uncertainties.

By separating scientific capabilities and workflow orchestration from the underlying foundation models, this architecture also allows newer models to be adopted without repeating domain-specific training and supports flexible deployment, including locally hosted open-weight models for projects with data-privacy or institutional requirements.

To address these challenges, we present the \textbf{M}ulti-agent \textbf{A}stronomy \textbf{A}ssistant for \textbf{S}cientific \textbf{A}nalysis and \textbf{I}nference (\maasai{}), a modular multi-agent system designed to support astronomical data analysis through natural-language interaction. \maasai{} combines foundation-model reasoning with scientific knowledge retrieval and specialised astronomical tools to plan and execute scientific workflows, ranging from individual analysis tasks to multi-step operations, while also supporting optional human-in-the-loop interaction and flexible deployment.

The development of \maasai{} is primarily motivated by the demands of next-generation radio surveys in the SKA era, with radio astronomy serving as its initial application domain. These surveys involve increasingly complex, multi-stage data-reduction and analysis workflows that require the integration of heterogeneous observational data, specialised software, and machine-learning models, thereby creating opportunities for agent-based assistance in their planning and execution. In this work, we therefore evaluate \maasai{} through representative radio-astronomy tasks, including quantitative benchmarks of image and source analysis, demonstrations of single- and multi-step workflows involving astronomical tools and archives, and a dedicated benchmark of literature-grounded scientific reasoning. At the same time, the framework is designed to support applications in other astronomical domains through additional tools and scientific knowledge resources.

The paper is organised as follows. Section~\ref{sec:literature-review} reviews related work on agentic AI in astronomy. Section~\ref{sec:method} describes the \maasai{} architecture and implementation. Section~\ref{sec:results} presents the evaluation results, while Section~\ref{sec:conclusions} discusses the main findings, limitations, and future developments.

\section{Related work}
\label{sec:literature-review}
Recent advances in LLMs have stimulated growing interest in agent-based and multi-agent AI systems that combine reasoning, planning, knowledge retrieval, and external tool execution to perform complex tasks \citep{Han2024,Wei2025,Qu2025,Krishnan2025,Zhou2026}. Unlike conventional conversational systems, these architectures coordinate specialised components with distinct responsibilities, such as planning, information retrieval, tool execution, and result evaluation. General frameworks such as \texttt{LangChain/LangGraph}\footnote{\url{https://docs.langchain.com}}, \texttt{CAMEL-AI}\footnote{\url{https://docs.camel-ai.org/}}, \texttt{AutoGen}\footnote{\url{https://microsoft.github.io/autogen}}, and \texttt{CrewAI}\footnote{\url{https://docs.crewai.com/}} provide mechanisms for coordinating such components through structured communication, shared state, or graph-based workflows.

Agentic AI systems have recently begun to emerge in astronomy, addressing different stages of scientific research. \emph{StarWhisper} \citep{StarWhisper} combines an LLM-based agent with specialised workflows and function calls to support telescope observation planning, control, and data processing through natural-language interaction. \emph{AstroReview} \citep{Wang2025} introduces a multi-agent framework for telescope-proposal evaluation, combining literature-based assessment of scientific novelty and merit, simulation-based feasibility analysis, and the synthesis of multiple reviews. In astrobiology, \emph{AstroAgents} \citep{Saeedi2025} coordinates specialised agents for mass-spectrometry data analysis, hypothesis generation, literature review, and scientific critique.

Other systems focus more directly on astronomical data analysis by combining LLM-based reasoning with specialised scientific software. \emph{Mephisto} \citep{Zechang2025} employs multiple agents to interpret multi-band galaxy observations through iterative reasoning and interaction with the \texttt{CIGALE} spectral-energy-distribution modelling software. \emph{Egent} \citep{egent} uses an LLM agent to inspect and iteratively refine Voigt-profile fits, supporting automated equivalent-width measurements and quality control in stellar spectra. In ground-based gamma-ray astronomy, \citet{Kostunin2025} developed LLM-based agents integrated with CTAO software and documentation, including an agent for generating and validating analysis code using \texttt{Gammapy}. A related, but methodologically different, approach was presented by \citet{SofiaAgent}, who used reinforcement learning to optimise source-finding parameters in \texttt{SoFiA-2} based on performance feedback from simulated SKA data. Together, these studies demonstrate different approaches to automating specialised astronomical analyses through interaction with established scientific software.

More comprehensive architectures have been developed to support multi-stage astrophysical research and data-analysis workflows. \emph{AI Cosmologist} \citep{Moss2025} coordinates specialised agents for planning, code generation and execution, result analysis, literature retrieval, and scientific synthesis. The system iteratively explores alternative approaches, evaluates their outcomes, and can extend the workflow to the preparation of research outputs. \emph{CmbAgent} \citep{cmbagent} introduces a multi-agent Planning \& Control architecture combining planning and coding agents with scientific literature retrieval, software-specific knowledge, local code execution, and iterative error correction. Its capabilities were demonstrated through an autonomous cosmological parameter-inference task using Type Ia supernova data, while the framework is designed to support quantitative scientific research more broadly. \emph{AstroGenesis} \citep{Sahakyan2026} presents a multi-agent framework for blazar research that integrates literature retrieval, multiwavelength observational data analysis, physical modelling, and research ideation. Its specialised agents coordinate multi-stage workflows, including broadband emission modelling using pretrained neural-network surrogate models.

\begin{figure*}[!htb]
\centering%
\includegraphics[scale=0.55]{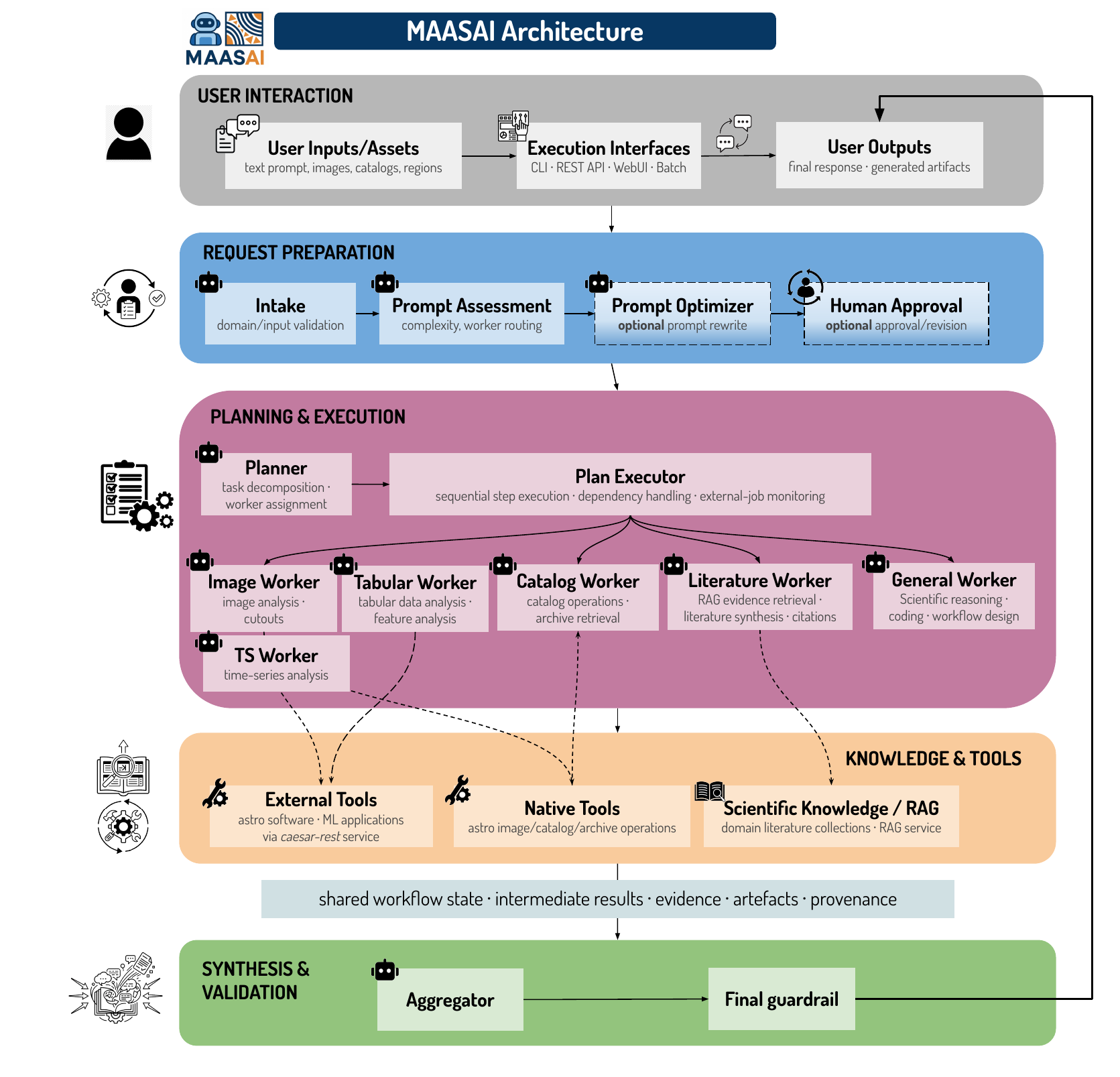}
\caption{High-level architecture of the \maasai{} system. User requests are processed through request-preparation, planning, execution, and synthesis stages, with specialised workers accessing scientific knowledge resources and astronomical tools. Dashed boxes indicate optional processing stages, including prompt optimisation and human approval.}%
\label{fig:maasai-architecture}
\end{figure*}

The application of CmbAgent to more open-ended research was subsequently investigated by \citet{Borrett2026}. In this work, the system was used to generate, execute, and iteratively refine cosmological inference pipelines for the \textit{FAIR Universe Weak Lensing Uncertainty Challenge}. Combining agent-driven exploration with human guidance, the approach achieved first place in the public phase and in one of the three final challenge leaderboards.

Beyond astronomy, more general agentic systems have been developed to support scientific discovery and research automation. \emph{Co-Scientist} \citep{Gottweis2026} coordinates specialised agents to generate, critique, rank, and refine scientific hypotheses using existing evidence. \emph{The AI Scientist} \citep{Lu2026} targets end-to-end automation of computational research, from idea generation and experiment execution to result analysis, manuscript preparation, and automated peer review. Similarly, \emph{Denario} \citep{denario2025} provides a modular multi-agent framework covering research ideation, literature-based novelty assessment, methodology design, code execution, scientific visualisation, manuscript preparation, and peer review across multiple disciplines, including astrophysics. Denario also supports CmbAgent as a backend for more extensive scientific analysis, illustrating how specialised research systems can be integrated into broader research-automation environments.

These developments span a wide range of objectives, from telescope operations and domain-specific data analysis to multi-stage scientific investigation and research automation. \maasai{} is situated within this emerging landscape as a modular multi-agent assistant focused on data and literature analysis, rather than on end-to-end automation of research idea generation and manuscript preparation. This reflects an emphasis on supporting researchers with the technical aspects of their work while preserving their responsibility for scientific direction and research outcomes. In this context, \maasai{} complements specialised astronomical analysis systems as well as broader AI-for-science frameworks targeting more extensive automation of the scientific research lifecycle. The following section describes its architecture and implementation.


\section{A multi-agent system for astronomical data analysis}
\label{sec:method}
\maasai{} is a multi-agent system for orchestrating scientific workflows in astronomy by combining large language models (LLMs), retrieval-augmented generation (RAG), multimodal reasoning, and specialised domain-specific analysis tools within a unified graph-based execution framework\footnote{The current implementation, publicly available at \url{https://github.com/inaf-oact-ai/maasai}, is based on the \texttt{LangGraph}/\texttt{LangChain} libraries. \texttt{LangGraph} (\url{https://docs.langchain.com/oss/python/langgraph/overview}) provides the stateful graph-based orchestration engine used to coordinate workflow execution, while \texttt{LangChain} (\url{https://docs.langchain.com/oss/python/langchain/overview}) supplies abstractions for language-model interaction, prompt management, agent execution, and tool integration.}. 

While initially developed for radio astronomy, the modular architecture of \maasai{} enables the progressive incorporation of other astronomical domains through dedicated scientific knowledge collections and domain-specific tools, without requiring changes to the overall orchestration framework. Initial developments in this direction include retrieval collections for solar physics and exoplanet science, alongside specialised analysis tools.

The following sections describe the principal architectural components of \maasai{}, including the graph-based orchestration workflow, retrieval-augmented scientific reasoning, the scientific tool ecosystem, and deployment infrastructure.

\subsection{System architecture}
\label{subsec:architecture}
\begin{figure}[!htb]
\centering%
\includegraphics[scale=0.4]{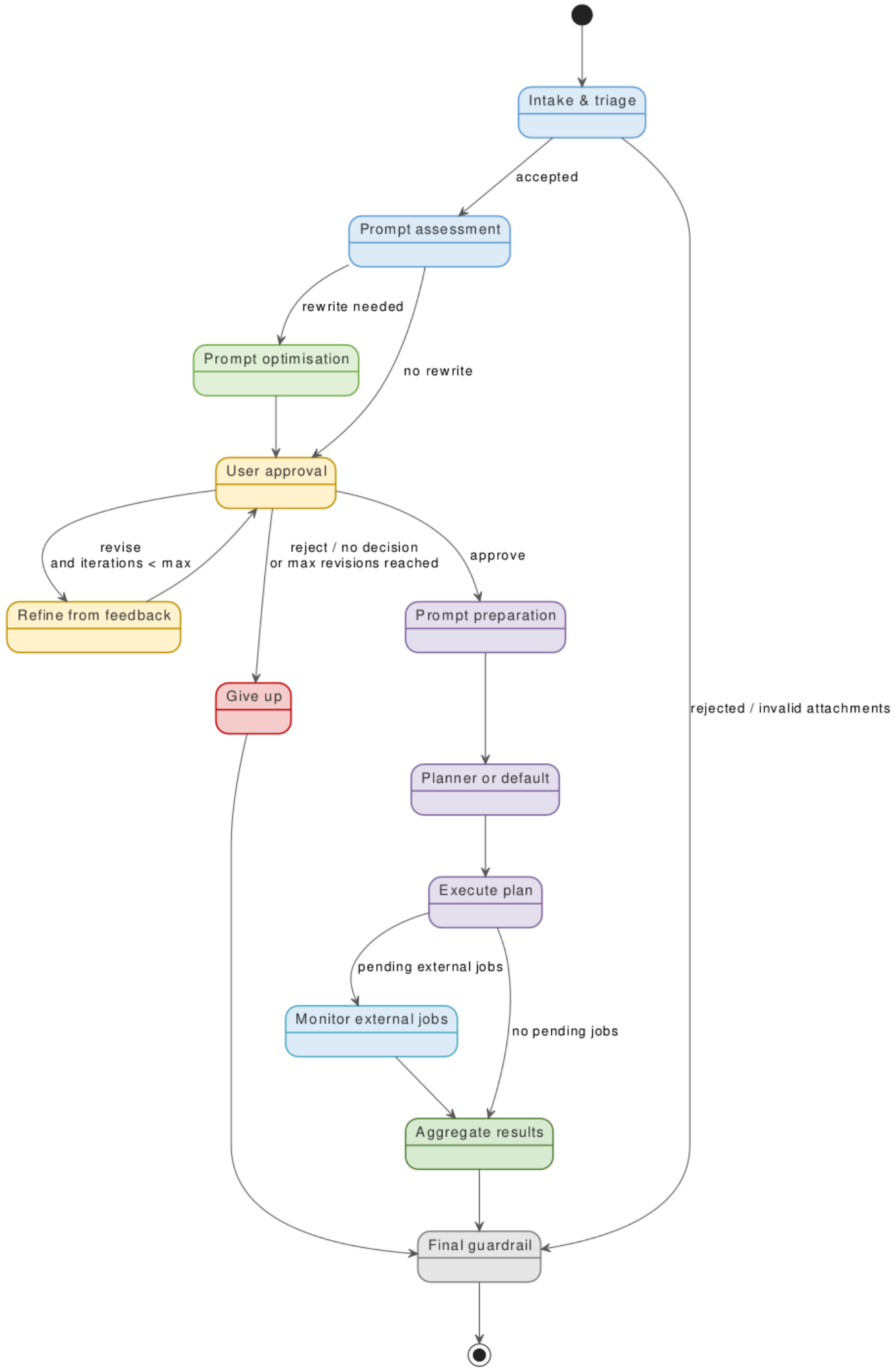}
\caption{\maasai{} workflow and execution logic. The diagram illustrates the main processing stages and conditional execution paths, from request intake and prompt assessment through prompt refinement, planning, execution, result aggregation, and final validation. The user-approval stage supports a human-in-the-loop refinement cycle when revision is requested, while external job monitoring is activated when asynchronous jobs are pending.}%
\label{fig:maasai-graph}
\end{figure}

The overall architecture of \maasai{}, illustrated in Fig.~\ref{fig:maasai-architecture}, is organised around specialised agents and workers coordinated through a stateful graph-based execution model. Workflow execution is internally represented as a stateful directed computational graph, whose nodes correspond to functional processing stages and whose edges define execution flow and conditional routing between agents and scientific tools. Fig.~\ref{fig:maasai-graph} provides a schematic representation of this workflow, highlighting the main processing stages, conditional execution paths, and the human-in-the-loop prompt refinement cycle.

The execution graph comprises successive processing stages responsible for prompt assessment, task planning and execution, and response aggregation. These components are described in the following subsections.

\subsubsection{Request triage and preparation}
\label{subsec:architecture-intake}
Incoming user requests, together with optional multimodal assets such as images or tabular datasets, are first processed by an \emph{intake} agent responsible for validating language constraints (English-only workflows), domain relevance (astronomy-related requests), and input consistency (e.g. astronomical image content). This triage stage can be assigned to a lightweight language model to reduce computational overhead and prevent unrelated or invalid requests from propagating to subsequent workflow stages.

Enabled requests are then processed by a \emph{prompt-assessment} agent that evaluates their complexity, identifies suggested worker types, and determines if they require internal rewriting, direct execution, or multi-step planning.
The resulting structured assessment is stored in the shared workflow state, and subsequently consumed by the planning and execution layers, enabling adaptive routing throughout the execution graph. Straightforward requests may therefore bypass explicit planner invocation and be dispatched through a default single-step plan, whereas more complex scientific analyses trigger planner-generated multi-step workflows.

When rewriting is required, a \emph{prompt-optimizer} agent uses the assessment output together with metadata extracted from attached assets to produce a clearer execution-ready request while preserving the user's scientific intent. The rewritten prompt records any explicit assumptions and unresolved questions separately from the execution request, allowing downstream stages to distinguish inferred context from user-provided information.

Before resource-intensive execution begins, \maasai{} provides an optional human-in-the-loop approval stage. The interface presents either the rewritten prompt or, when no rewriting is required, the original request, together with the inferred assumptions, unresolved ambiguities, execution rationale, estimated complexity, and proposed execution strategy. Users may approve the request, request further revisions, or reject the workflow before planning and execution proceed.

\subsubsection{Task planning}
\label{subsec:task-planning}
Following request preparation, the planning layer transforms high-level scientific requests into executable workflows. Based on the outcome of the prompt-assessment stage, \maasai{} either generates a default single-step execution plan or invokes the planner agent to decompose the request into a structured multi-step workflow.

For complex analyses, the planner uses the prepared execution prompt, structured prompt assessment, metadata describing attached multimodal assets, and, when enabled, retrieval-augmented scientific context to construct an execution plan. The plan specifies the execution steps, expected outputs, required worker types, input dependencies, execution constraints, and rationale. Worker assignment reflects the scientific requirements of each step, distinguishing, for example, between image analysis, catalogue and archive operations, numerical feature-space analysis, time-series processing, and literature-supported reasoning.

The execution plan is then passed to the execution layer, which coordinates worker execution, manages inter-step dependencies, monitors execution outcomes, and orchestrates the resulting workflow.

\subsubsection{Task execution}
\label{subsec:task-execution}
Task execution is coordinated through an orchestration layer that routes each execution step to the specialised worker agent assigned during the planning stage. The current implementation defines seven worker classes: \emph{image}, \emph{catalog}, \emph{tabular}, \emph{timeseries}, \emph{measurement-set}, \emph{literature}, and \emph{general}. Each worker operates as a specialised agent with access to a dedicated set of tools, retrieval capabilities when applicable, and execution constraints.


The separation of workers reflects the different semantics of the scientific data products they handle:

\begin{itemize}
\item The \emph{catalog} worker processes astronomical source catalogues and coordinates, including positional associations and archive interactions.
\item The \emph{tabular} worker operates on numerical feature matrices, supporting statistical analysis, dimensionality reduction, clustering, and anomaly detection.
\item The \emph{timeseries} worker handles temporally ordered observations, including light curves, where sampling information and temporal structure must be preserved.
\item The \emph{measurement-set} worker handles radio-interferometric visibility datasets and associated processing operations.
\end{itemize}

Execution proceeds sequentially over the ordered task plan produced by the planning layer. For each execution step, the orchestration layer constructs a worker-specific execution prompt containing the approved user request, the current execution-step description, available multimodal assets, structured outputs and artefacts produced by previously completed steps, and retrieved scientific context when available. Intermediate products are propagated through the shared workflow state and selectively forwarded to downstream workers according to explicit step dependencies, allowing later stages to consume results generated by earlier analyses.

Scientific operations are performed through native tools or external applications accessed via \texttt{caesar-rest}\footnote{\url{https://github.com/inaf-oact-ai/caesar-rest}}, a REST-based service for managing the execution of containerised astronomical applications on computing clusters \citep{Riggi2021}(Section~\ref{subsec:tools}). Both classes of tools expose structured interfaces and return results and artefact references in a common representation, allowing scientific products to be exchanged between workflow stages.

When an external scientific application is selected, execution is delegated to the \texttt{caesar-rest} service infrastructure. Submitted jobs are represented by structured execution metadata and monitored until completion by periodically polling the backend service. Output products are then retrieved, normalised into the common workflow representation, registered as workflow artefacts, and injected into the shared workflow state. Consequently, native and external tools expose a consistent execution interface despite their different execution backends.

Following execution, results produced by individual workers are consolidated through a dedicated aggregation stage that combines structured outputs, generated artefacts, retrieved evidence, execution metadata, provenance information, and external-tool results into a coherent scientific response. During this stage, citations are normalised, execution caveats and partial failures are propagated, and a final validation step verifies consistency with workflow policies before the response is returned to the user.

\begin{table*}[htbp]
\centering
\footnotesize%
\caption{Example of machine-readable application metadata dynamically
discovered from the \texttt{caesar-rest} backend for the
\texttt{caesar-yolo} application. The table shows an abridged view of the
metadata made available to \maasai{} for tool selection, configuration,
execution, and interpretation of returned products.}
\label{tab:caesar-yolo}
\begin{tabular}{p{3.2cm} p{11.5cm}}
\hline
\textbf{Metadata field} & \textbf{Example content} \\
\hline

{\scriptsize Application} &
{\scriptsize \texttt{caesar-yolo}.} \\

{\scriptsize Scientific description} &
{\scriptsize Application of a pretrained YOLO object-detection model to
radio-astronomical images for source detection and morphological
classification into \texttt{spurious}, \texttt{compact},
\texttt{extended}, \texttt{extended-multisland}, and \texttt{flagged}
classes.} \\

{\scriptsize Input requirements} &
{\scriptsize Single radio-continuum astronomical image; supported formats:
FITS and PNG.} \\

{\scriptsize Configurable options} &
{\scriptsize Structured parameter definitions specifying name, type,
description, category, default value, and, where applicable, allowed values
or numerical ranges. Options include pretrained model selection, image
preprocessing, detection thresholds, image tiling and parallel-processing
parameters, and output configuration.} \\

{\scriptsize Primary output} &
{\scriptsize Required JSON source catalogue
(\texttt{role=primary\_result}, \texttt{parser=json}) containing detected
sources with class identifier and label, confidence score, bounding-box
coordinates, and detection flags.} \\

{\scriptsize Additional outputs} &
{\scriptsize Optional PNG detection visualisation
(\texttt{role=visualization}), DS9 region file
(\texttt{role=visualization}), and execution log
(\texttt{role=diagnostic}), each with declared file type and parser.} \\

{\scriptsize Output contract} &
{\scriptsize Machine-readable descriptions of generated products,
including filename patterns, MIME types, semantic roles, parsers,
required/optional status, and format information. For the primary catalogue,
the schema also describes the JSON structure and individual source fields.} \\

{\scriptsize Declared limitations} &
{\scriptsize Detection accuracy depends on the selected pretrained model,
preprocessing, input-image survey characteristics, and detection
parameters. The recommended image size should match the training
configuration of the selected model; very large images require tiling and
parallel execution. Models are trained and tested on radio images, and
performance in other astronomical domains may therefore be sub-optimal.} \\

\hline
\end{tabular}
\end{table*}

\subsection{Scientific knowledge retrieval}
\label{subsec:rag}
\maasai{} integrates an optional retrieval-augmented generation (RAG) subsystem that provides literature-grounded scientific context during workflow execution\footnote{The retrieval subsystem, publicly available at \url{https://github.com/inaf-oact-ai/llama-index-rag}, is implemented using \texttt{LlamaIndex}, \texttt{Qdrant}, and embedding-based semantic search. It supports both local and service-oriented deployment through a \texttt{FastAPI}-based retrieval service exposing REST APIs for semantic search and multi-collection querying, together with a dedicated \texttt{Streamlit} web interface for interactive exploration and validation.}. The knowledge base is organised into multiple vector collections covering different astrophysical domains, currently including radio astronomy, solar physics, and exoplanet science, and contains scientific papers, review articles, books, and other curated scientific resources.

The retrieval subsystem adopts a modular architecture in which semantically indexed collections are queried according to the current task. Rather than relying on a single monolithic vector store, retrieval requests may be guided by domain hints derived from workflow assessment or by the worker currently executing the task. This organisation allows new scientific domains to be incorporated without modifying the higher-level orchestration framework.

RAG retrieval can be invoked at two different stages of the workflow. During planning, an optional retrieval step may provide high-level scientific context that assists the planner in constructing an appropriate execution strategy. This planner context is used solely to improve workflow decomposition and worker assignment and is not treated as scientific evidence for the final response. During execution, the specialised literature worker performs an independent semantic retrieval using the approved execution request together with the current plan step, obtaining passages that constitute the evidence base for literature-grounded responses and citation generation. When no dedicated literature step is present, planner retrieval context may optionally be propagated to downstream reasoning workers as supplementary background information.

Retrieved documents are represented internally as structured objects containing document identifiers, similarity scores, collection names, bibliographic metadata, and retrieved text passages. These evidence objects are propagated through the workflow state together with execution results, allowing downstream stages to distinguish retrieved scientific evidence from language-model reasoning and tool-generated outputs. Final response synthesis is therefore performed by the language model while maintaining explicit provenance for literature-derived statements.

\begin{figure*}[!htb]
\centering%
\includegraphics[scale=0.55]{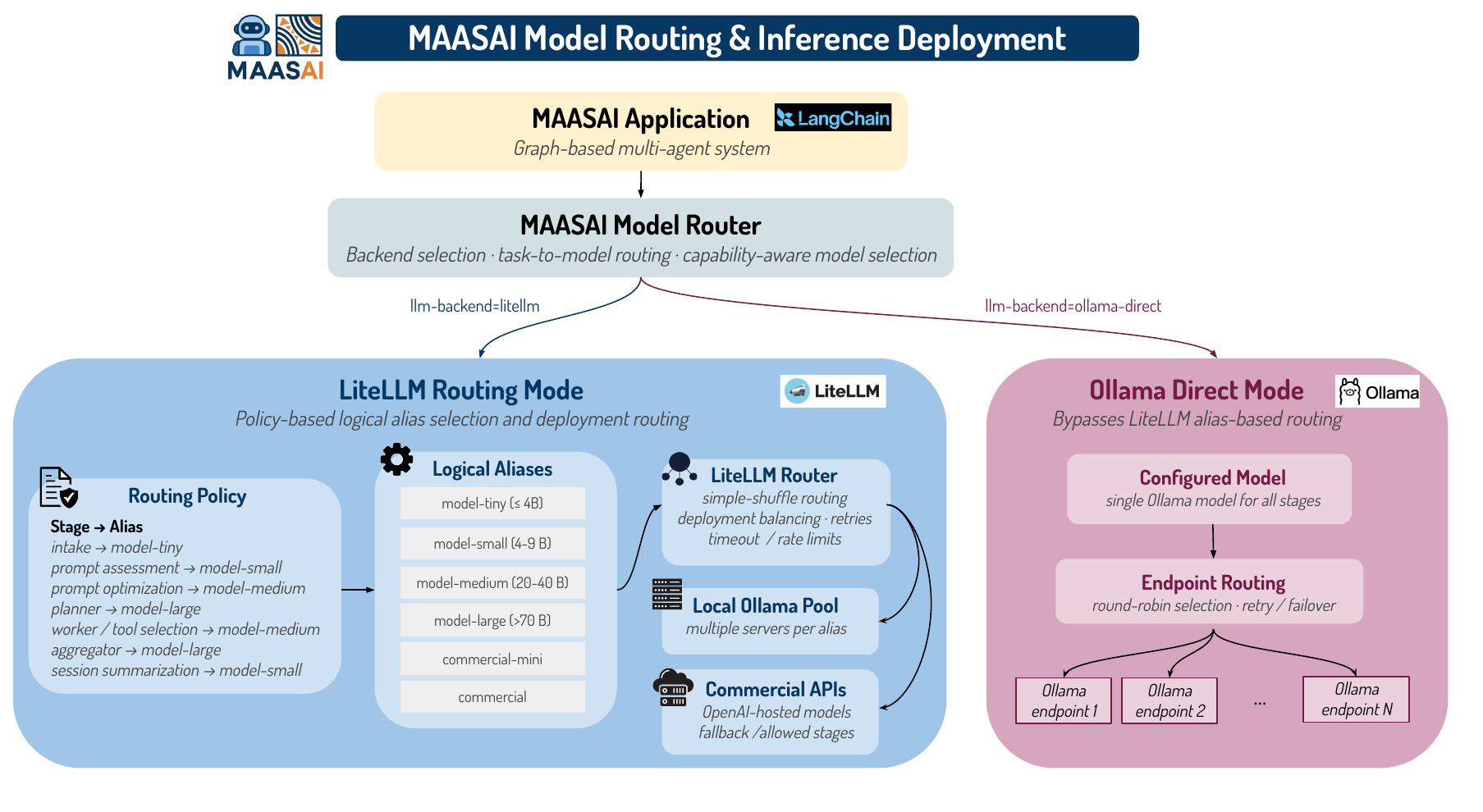}
\caption{\maasai{} model routing and inference deployment architecture. The \maasai{} model router supports two inference modes. In \textit{Routing Mode}, workflow stages are mapped to logical model aliases according to a configurable routing policy; \texttt{LiteLLM} then routes requests among local \texttt{Ollama} deployments or commercial API models associated with each alias, providing deployment-level balancing, retries, timeouts, and rate limiting. In \textit{Direct Mode}, \texttt{LiteLLM} alias routing is bypassed and a single configured \texttt{Ollama} model is used across workflow stages; the router cycles across one or more configured \texttt{Ollama} endpoints, while router-mediated invocations can retry on subsequent endpoints following failures. This separation allows the same multi-agent application to operate with heterogeneous model deployments or with a simpler fully local inference configuration.}%
\label{fig:maasai-deployment}
\end{figure*}

\subsection{Astronomical tools and external scientific services}
\label{subsec:tools}

Scientific analyses in \maasai{} are performed through an extensible tool framework that delegates specialised operations to scientific software rather than implementing them within the language-model reasoning process.

The current implementation includes two classes of tools. Native tools are implemented within the \maasai{} codebase and provide operations for FITS image inspection, image cutout generation, source measurement, source-catalog manipulation, visualisation, archive queries, archive cross-matching, and catalogue enhancement. External tools integrate existing astronomical software and machine-learning applications through the \texttt{caesar-rest} service API \citep{Riggi2021}, including traditional radio-source extraction software (e.g. \texttt{caesar}, \citealt{Riggi2016,Riggi2019}, and \texttt{Aegean}, \citealt{Aegean,Aegean2}), deep-learning object detectors (e.g. \texttt{caesar-yolo}, \citealt{Riggi2023}), image and source classifiers, feature-extraction pipelines, similarity-search methods, clustering algorithms, and dimensionality-reduction techniques. The current tool inventory is summarised in Table~\ref{tab:maasai_tools}.

Tools expose standardised interfaces and are dynamically discovered and registered at runtime through a unified registry. Native tools are registered from the \maasai{} codebase, whereas external applications are discovered from the \texttt{caesar-rest} backend and exposed as \texttt{LangChain}-compatible tools. Their machine-readable metadata describe scientific capabilities, input requirements, configurable parameters, expected output products, and operational limitations (see Table~\ref{tab:caesar-yolo} for an example). Tool results, including generated artefacts, execution metadata, and provenance information, are represented in a common format that enables data exchange between workflow stages.

Native tools typically perform lightweight operations within the \maasai{} execution environment, whereas external applications can support more computationally demanding analyses through the \texttt{caesar-rest} infrastructure, which enables the execution of containerised scientific applications on a \texttt{Slurm}-managed computing cluster. The \maasai{} orchestration layer submits external execution requests, monitors asynchronous jobs, retrieves generated products, and registers them as workflow artefacts for subsequent processing. This common interface allows native and external scientific tools to be combined within multi-step workflows.

\begin{figure*}[!htb]
\centering%
\subtable[\{\texttt{EXTENDED,DIFFUSE}\}]{\includegraphics[scale=0.25]{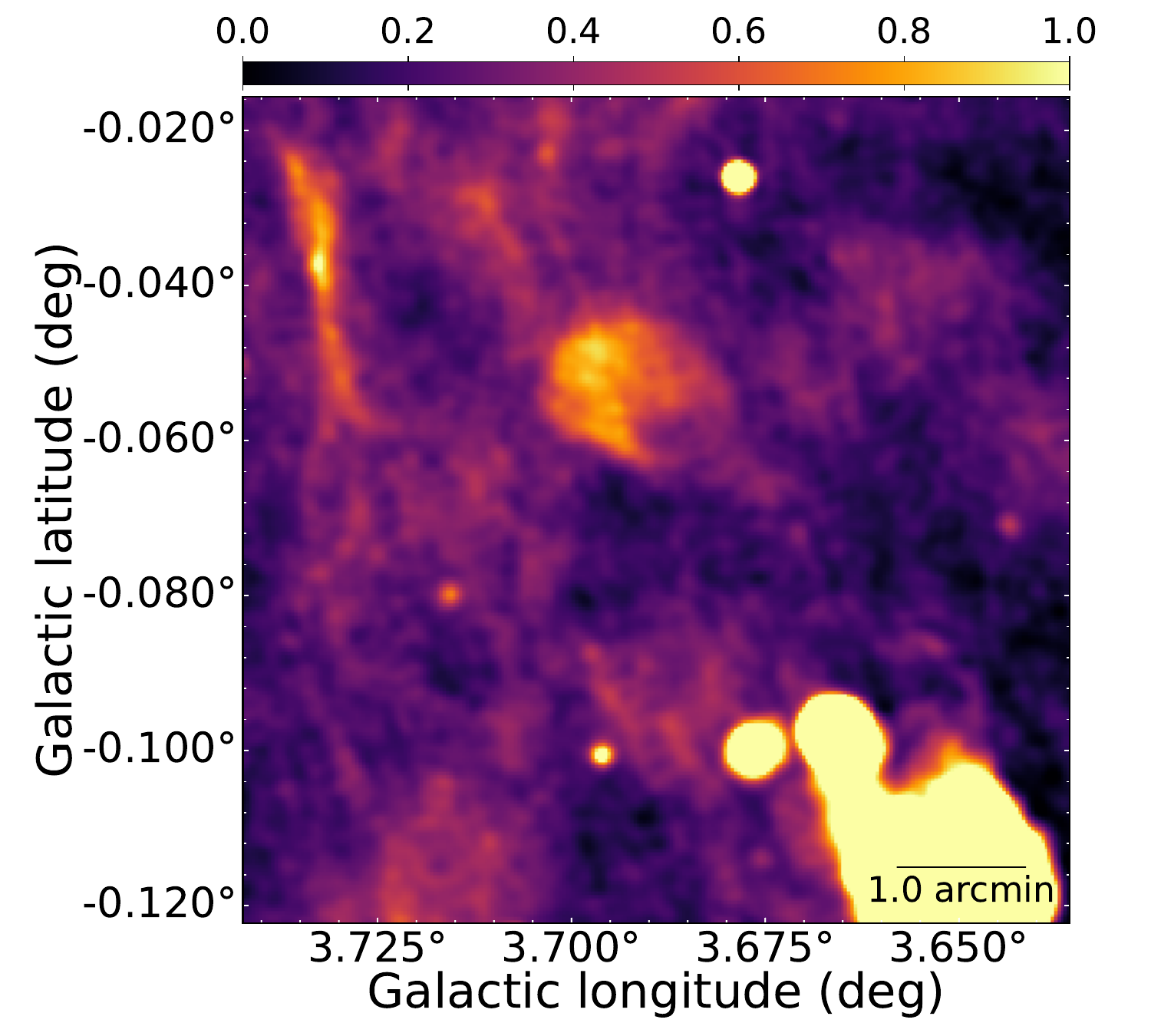}}%
\subtable[\{\texttt{DIFFUSE-LARGE}\}]{\includegraphics[scale=0.25]{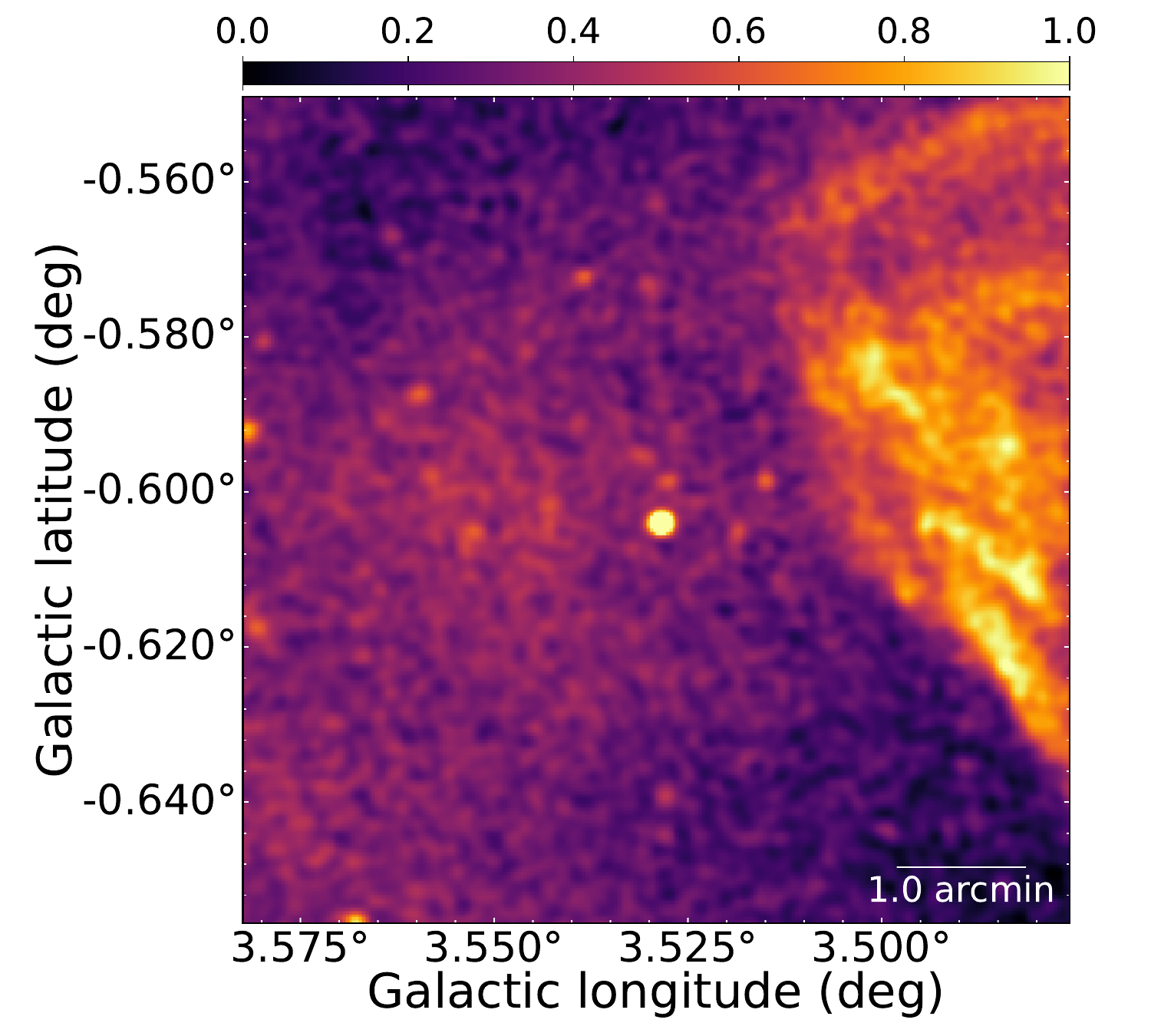}}\\%
\vspace{-0.1cm}%
\subtable[\texttt{2C-2P}]{\includegraphics[scale=0.25]{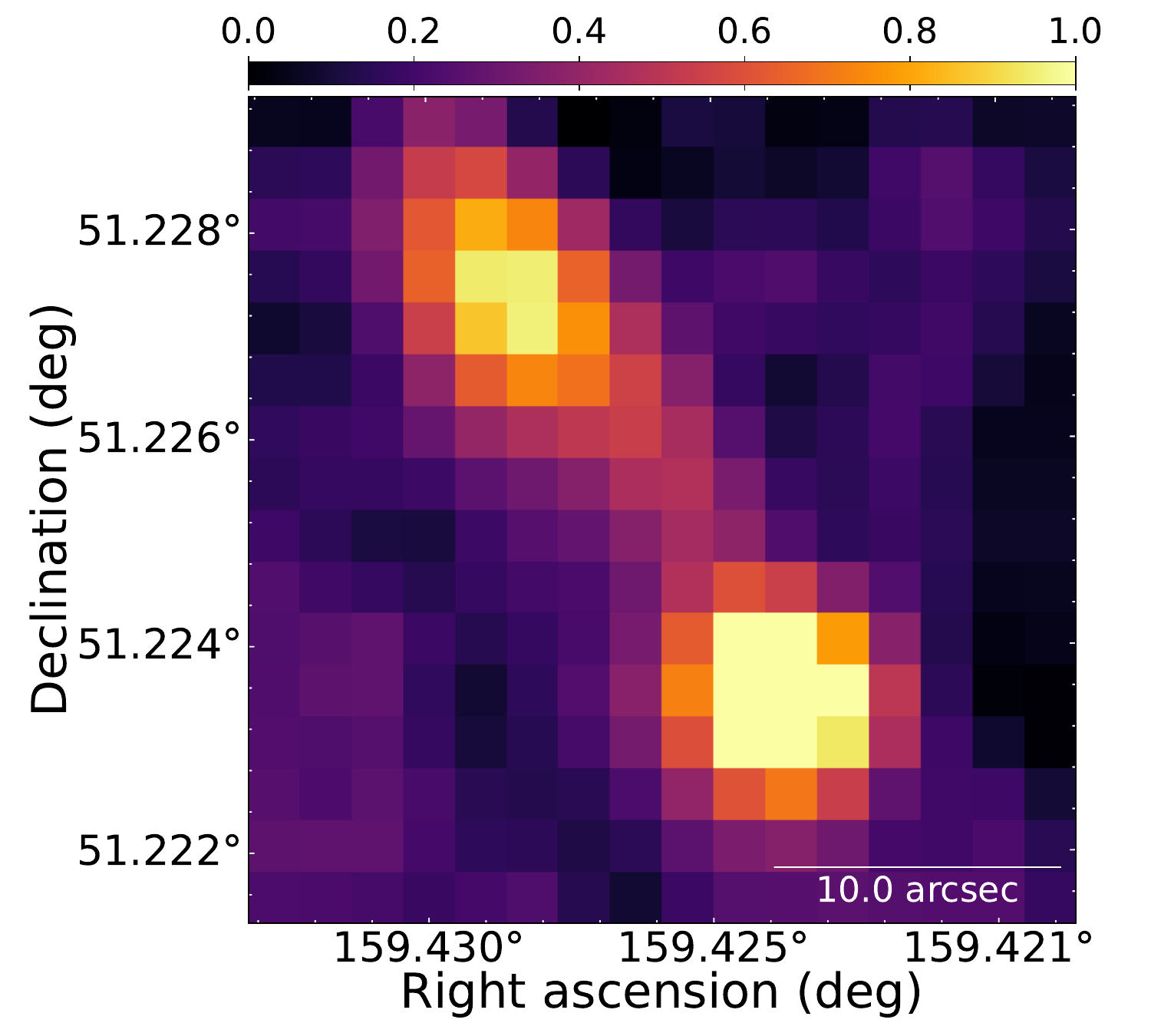}}%
\subtable[\texttt{3C-3P}]{\includegraphics[scale=0.25]{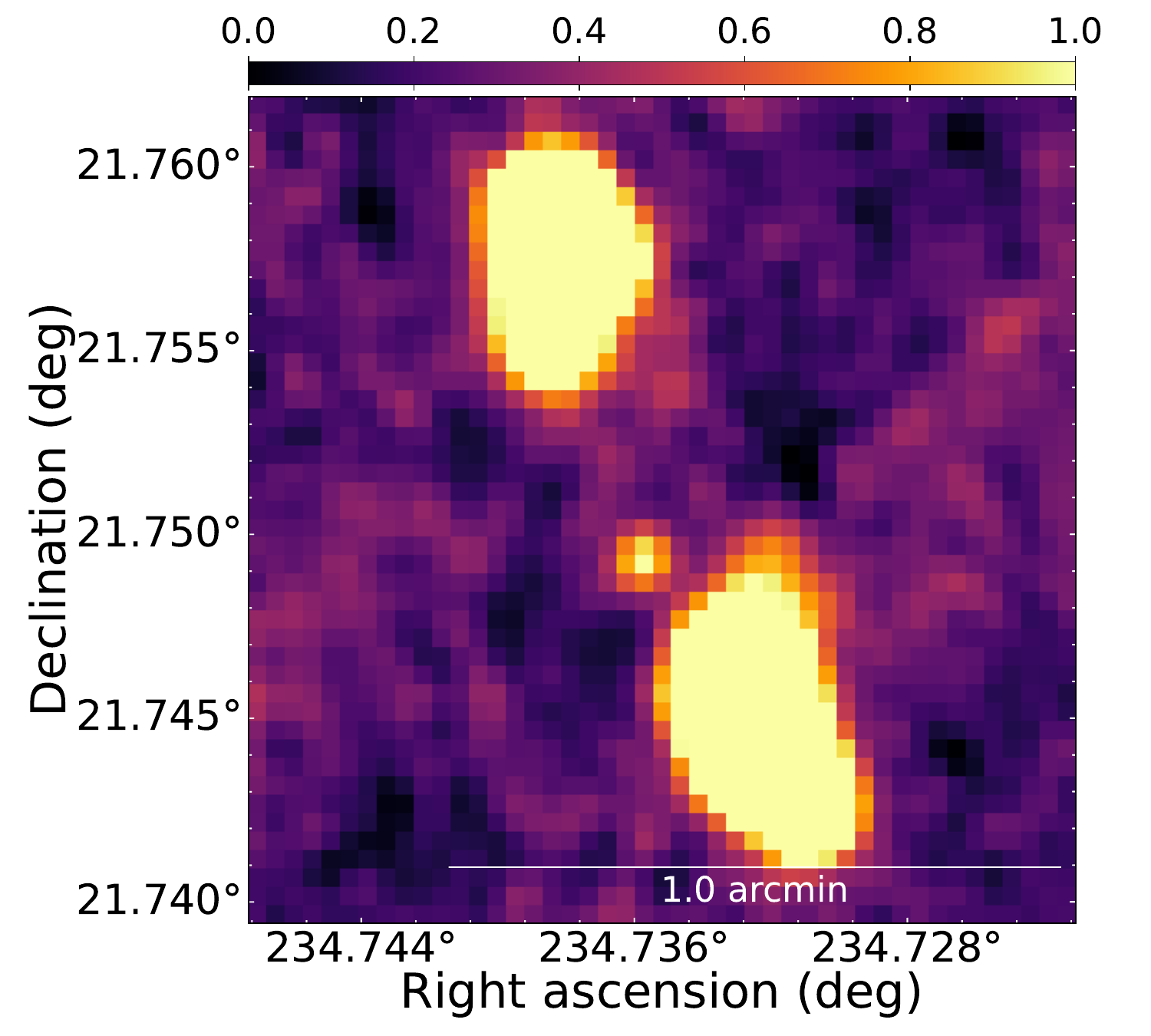}}%
\caption{Sample annotated radio images for the B1 (top panel) and B2 (bottom panel) radio benchmarks. Annotation class labels are reported below each image. Data are taken from the MeerKAT SMGPS \citep{SMGPS} and VLA FIRST \citep{FIRST} surveys for B1 and B2 benchmarks, respectively.}%
\label{fig:sampledata_B1_B2}
\end{figure*}

\subsection{Configuration and deployment}
\label{subsec:configuration}
Fig.~\ref{fig:maasai-deployment} summarises the configuration and model-routing strategy adopted in \maasai{}. The framework separates workflow logic from the underlying inference and execution infrastructure through a hierarchical configuration system. Runtime behaviour is controlled through environment variables defining deployment defaults, \texttt{LiteLLM} configuration files specifying model aliases and routing policies, and command-line overrides for individual execution parameters.

\maasai{} supports two inference modes. In \texttt{LiteLLM} routing mode, workflow stages are associated with logical model classes (e.g. \texttt{model-tiny}, \texttt{model-small}, \texttt{model-medium}, and \texttt{model-large}), mapped to local or remote inference backends through configurable routing policies. This allows lightweight stages, such as intake validation and prompt assessment, to use smaller models, while more demanding operations, such as planning, literature-grounded reasoning, and response aggregation, can employ more capable models. In direct mode, \texttt{LiteLLM} routing is bypassed and a single configured model is used across workflow stages, with support for multiple local inference endpoints.

Additional configuration parameters control workflow behaviour, including timeouts, prompt approval and rewriting, retrieval-augmented generation, scientific tool integration, artefact management, debugging, and execution monitoring. They also govern communication with external services, including the retrieval subsystem and the \texttt{caesar-rest} execution backend.

Workflow state and conversational context are managed through configurable \texttt{LangGraph} checkpoint backends supporting in-memory, \texttt{SQLite}, and \texttt{PostgreSQL} storage. Optional conversation summarisation and message trimming strategies help control context growth during long-running interactive sessions.

This configuration flexibility allows \maasai{} to operate with local open-weight models, commercial inference services, or heterogeneous model deployments, depending on computational resources, institutional requirements, and data-governance constraints.

\begin{figure*}[!htb]
\centering%
\subtable[\texttt{B1}]{\includegraphics[scale=0.4]{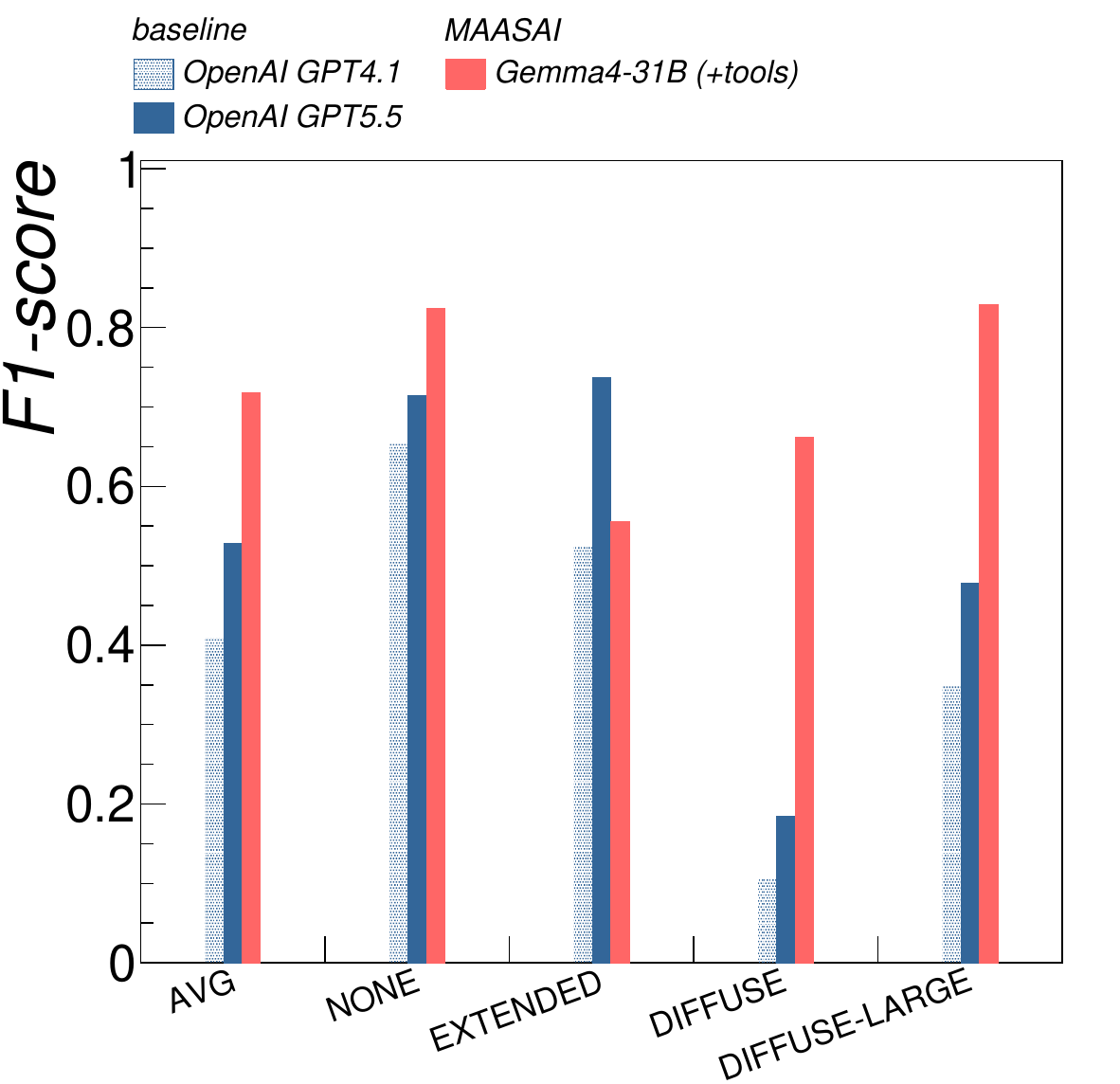}}%
\subtable[\texttt{B2}]{\includegraphics[scale=0.4]{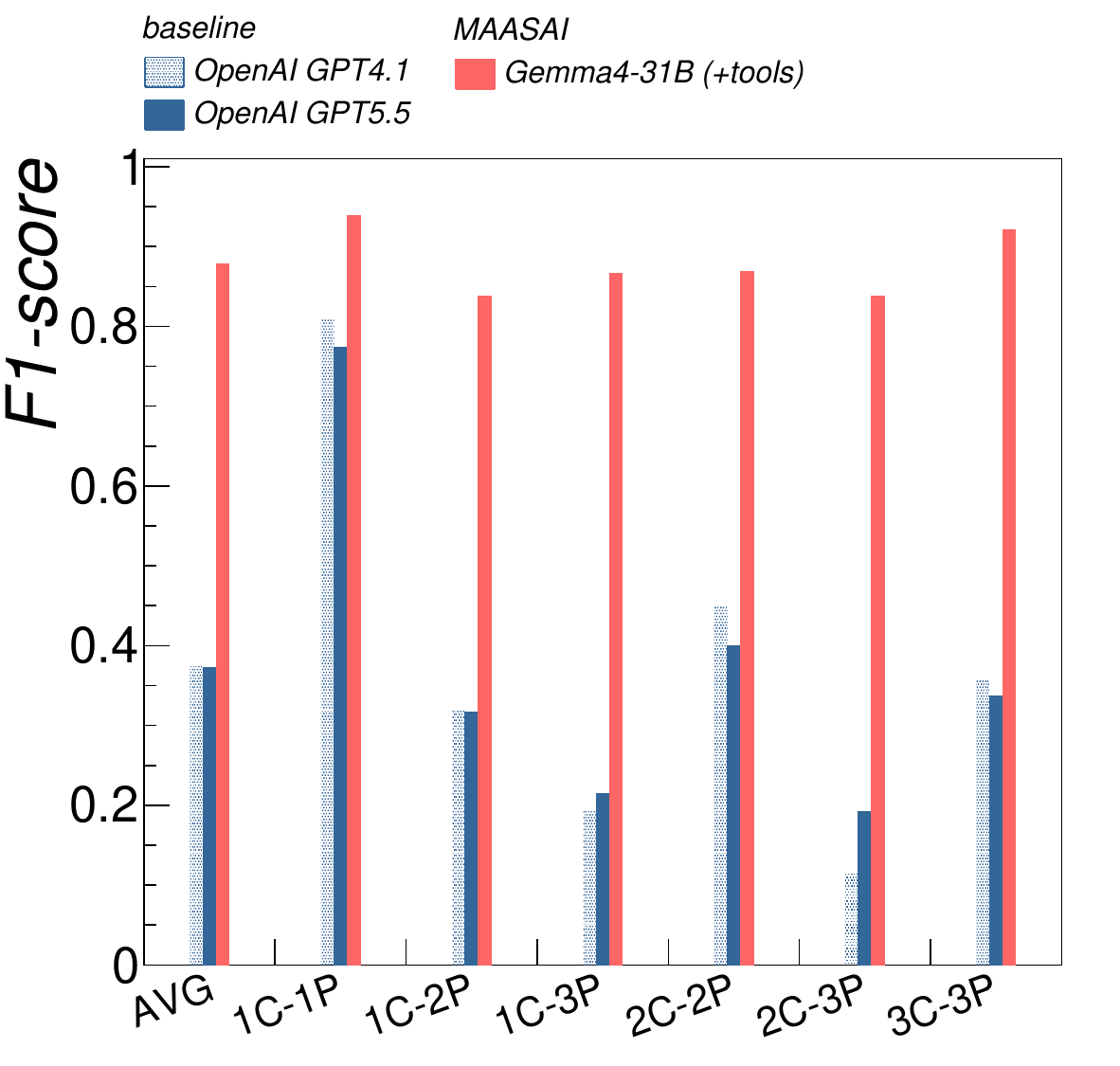}}%
\vspace{-0.4cm}%
\caption{Classification F1-scores (macro-averaged and per class) obtained on the B1 (left; extended/diffuse source detection) and B2 (right; source morphology classification) radio benchmarks. Results obtained with the \maasai{} system, configured with a Gemma4:31b model, are shown as filled red histograms, while those obtained with an OpenAI GPT baseline are shown with blue histograms: GPT5.5 (this work), GPT4.1 (taken from \citealt{Riggi2025}).}%
\label{fig:f1scores_B1_B2}
\end{figure*}

\subsection{Execution interfaces and data management}
\label{subsec:interfaces}
\maasai{} exposes its orchestration workflow through multiple execution interfaces, supporting interactive use, programmatic integration, and large-scale automated experiments without modifying the underlying execution graph. Three execution modes are currently supported:

\begin{itemize}
\item \emph{Command-line interface (CLI)} for interactive execution and rapid experimentation with natural-language prompts and optional scientific data input products;

\item \emph{REST API}, implemented using \texttt{FastAPI}, providing synchronous and streaming endpoints suitable for integration into external applications and scientific workflows;

\item \emph{Batch mode}, designed for benchmark evaluation and large-scale processing, supporting execution resumption, retry policies, and structured result collection.
\end{itemize}

A web interface is provided through an \texttt{OpenWebUI} integration that communicates with the REST service, enabling conversational interaction and real-time progress reporting for long-running workflows. For web and API deployments, \maasai{} supports authenticated user access through trusted identity headers or OAuth2/OpenID Connect tokens, allowing workflow sessions and generated artefacts to be associated with individual users. Integration with the \texttt{caesar-rest} backend additionally supports the propagation of authenticated user identity, allowing external scientific operations to be executed on behalf of the corresponding user.

Scientific inputs are handled through a unified attachment abstraction that automatically recognises common astronomical data products, including FITS images, source catalogues, and DS9 region files. Image products are converted into multimodal representations suitable for vision-language models while preserving references to the original scientific files for downstream analysis. Generated outputs are registered as persistent artefacts, allowing scientific products to be accessed and retrieved through the execution interfaces.

\section{Radio data analysis examples}
\label{sec:results}
To evaluate the capabilities of \maasai{}, we consider three representative categories of radio-astronomy tasks: (i) single-step image- or catalogue-based analysis tasks, (ii) multi-step scientific workflows requiring the orchestration of multiple tools, and (iii) literature-based scientific question answering using retrieval-augmented generation.

For these experiments, \maasai{} was configured to use a single open-weight multimodal model capable of tool calling across the astronomical tool suite described in Section~\ref{subsec:tools}. We explored Qwen3.5/3.6 and Gemma4 models, ultimately selecting Gemma4:31b because of its greater execution stability. The model was served locally through \texttt{Ollama}\footnote{\texttt{Ollama} is a platform for running and serving language models, supporting both local and cloud-based inference (\url{https://ollama.com}).} on a single NVIDIA GPU with 48~GB of memory. The tasks were also evaluated using OpenAI GPT-5.5 as a general-purpose commercial-model baseline, with identical input data and task instructions, enabling quantitative comparisons where ground truth was available and qualitative comparisons for the capability demonstrations.

\subsection{Single-step analysis tasks}
\label{subsec:single-step-analysis-tasks}
We first consider two image-classification benchmarks designed to assess the ability of \maasai{} to select and invoke appropriate scientific tools within a single execution step.

\subsubsection{B1 - Extended/diffuse source detection}
\label{subsec:benchmark_B1}
Benchmark B1 evaluates multi-label classification of radio images according to the presence of extended sources (\texttt{EXTENDED}), small diffuse emission (\texttt{DIFFUSE}), and large diffuse structures (\texttt{DIFFUSE-LARGE}). We used the \emph{radioimg-multilabel} test dataset described in \cite{Riggi2025}, containing 5,718 annotated images of size 256$\times$256 pixels from various radio surveys. Sample annotated images are shown in Fig.~\ref{fig:sampledata_B1_B2} (top panels). The complete task prompt is reported in \ref{app:benchmark-b1-prompt}.

Fig.~\ref{fig:f1scores_B1_B2} (left panel) reports the macro-averaged and per-class F1 scores, while the corresponding one-vs-rest confusion matrices are shown in the Appendix (Fig.~\ref{fig:cm_B1}).

The two approaches exhibit markedly different classification performance. \maasai{} achieves a higher macro-averaged F1 score than GPT-5.5 (0.72 vs 0.53), with substantial improvements for \texttt{NONE} (0.82 vs 0.71), \texttt{DIFFUSE} (0.66 vs 0.18), and \texttt{DIFFUSE-LARGE} (0.83 vs 0.48). Conversely, GPT-5.5 performs better on the \texttt{EXTENDED} class (F1=0.74 vs 0.56), primarily due to its higher recall (0.74 vs 0.41). For \texttt{DIFFUSE-LARGE}, GPT-5.5 achieves higher recall (0.95 vs 0.83) but substantially lower precision (0.32 vs 0.83), indicating a greater tendency to produce false positives. Overall, the tool-augmented \maasai{} system provides more reliable classification of diffuse emission, while the identification of extended sources remains a relative weakness.

\subsubsection{B2 - Source morphology classification}
\label{subsec:benchmark_B2}
Benchmark B2 evaluates single-label radio morphology classification, assigning each image to one of six classes defined by the number of disjoint radio islands and emission peaks. We used the \texttt{rgz-smorph} test dataset described in \cite{Riggi2025}, containing 3,835 radio images from the VLA FIRST survey, each centred on a radio source belonging to one of the six morphological classes: \texttt{1C-1P}, \texttt{1C-2P}, \texttt{1C-3P}, \texttt{2C-2P}, \texttt{2C-3P}, and \texttt{3C-3P}. Sample annotated images are shown in Fig.~\ref{fig:sampledata_B1_B2} (bottom panels). The complete task prompt is reported in \ref{app:benchmark-b2-prompt}.

The resulting F1 scores are shown in Fig.~\ref{fig:f1scores_B1_B2} (right panel), while the corresponding normalised confusion matrices are reported in the Appendix (Fig.~\ref{fig:cm_B2}). \maasai{} substantially outperforms GPT-5.5 across all six morphological classes, increasing the macro-averaged F1 score from 0.37 to 0.88. The improvement is particularly pronounced for complex morphologies, with F1 scores increasing from 0.21 to 0.87 for \texttt{1C-3P}, from 0.19 to 0.84 for \texttt{2C-3P}, and from 0.34 to 0.92 for \texttt{3C-3P}. For the simplest class, \texttt{1C-1P}, the F1 score increases from 0.77 to 0.94.

The confusion matrices show that GPT-5.5 frequently misclassifies sources with multiple components or emission peaks, particularly confusing \texttt{1C-2P} with \texttt{2C-2P} and assigning several complex morphologies to the \texttt{2C-2P} class. In contrast, \maasai{} achieves recalls of at least 83\% across all classes, with the remaining errors occurring mainly between morphologically related categories. Overall, these results highlight the advantage of tool-augmented analysis for fine-grained radio morphology classification, particularly for complex multi-component sources.

\begin{table*}[!ht]
\centering
\caption{Summary of representative single-step native-tool demonstrations. Unlike benchmarks~B1 and~B2, these experiments are intended as capability demonstrations rather than quantitative benchmarks.}
\label{tab:single_step_native_tools}
\footnotesize%
\begin{tabular}{p{2.5cm} p{4.4cm} p{4.3cm} p{5.5cm}}
\hline
\textbf{Scientific task} &
\textbf{Native capability} &
\textbf{Typical output} &
\textbf{Comparison with GPT-5.5} \\
\hline

Source measurements &
Aperture photometry from DS9 regions with local background estimation &
Flux densities, uncertainties, local-background statistics, morphology labels, and downloadable JSON/CSV catalogues &
Produced quantitatively consistent photometry. Morphology was inferred by GPT-5.5 using geometric heuristics, whereas \maasai{} preserved the input region annotations and exposed their provenance.\\

\hline%

Source cutout archive retrieval &
Multi-survey retrieval of archival FITS cutouts from astronomical archives using sky regions &
Survey-specific FITS cutouts, JSON retrieval manifest, downloadable archive, and retrieval-status report &
Correctly interpreted the regions and attempted all requested queries, but produced no FITS cutouts because of execution-environment limitations; \maasai{} retrieved and packaged the available WISE and 2MASS products while explicitly reporting failed or unsupported requests.\\

\hline
\end{tabular}
\end{table*}

\begin{figure}[!htb]
\centering%
\includegraphics[scale=0.35]{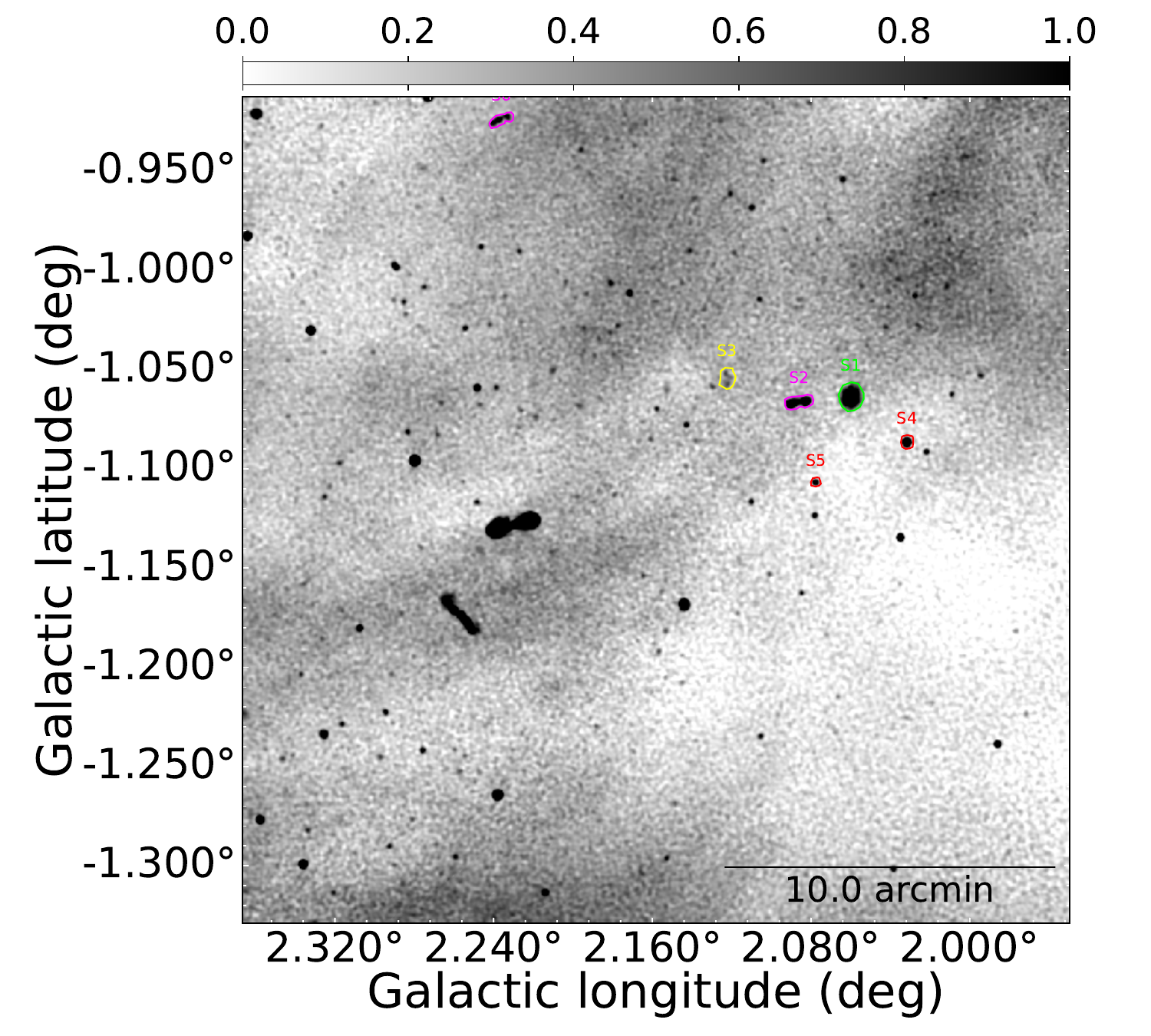}
\caption{Sample radio image, taken from the MeerKAT SMGPS survey \citep{SMGPS}, with selected source regions of different morphologies superimposed, used for testing \maasai{} astronomical source measurement and archive cutout tools.}%
\label{fig:sample-img-regions}
\end{figure}

\subsubsection{Additional single-step tool demonstrations}
\label{subsec:single-step-native-tasks}
Beyond the quantitative benchmarks, we evaluated representative native astronomical tools through single-step scientific requests focusing on source measurements and archival data retrieval. These experiments serve as capability demonstrations rather than formal benchmarks, and the comparison with GPT-5.5 is therefore qualitative. Table~\ref{tab:single_step_native_tools} summarises the evaluated capabilities, their typical scientific outputs, and the qualitative comparison with GPT-5.5.

\paragraph{Source measurements}\leavevmode%
\label{subsec:source-measurement-task}
We evaluated source-measurement capabilities using a request for aperture photometry of sources defined by DS9 regions, including local-background estimation, flux-density uncertainties, and morphology classification. For this test, we used the radio image and source regions shown in Fig.~\ref{fig:sample-img-regions}. The complete task prompt is reported in \ref{app:source-measurement-prompt}.

Both \maasai{} and GPT-5.5 successfully completed the analysis, producing downloadable structured outputs (JSON and CSV) together with human-readable summaries. GPT-5.5 performed the task by dynamically generating and executing Python code using established astronomical libraries, whereas \maasai{} delegated the computation to a validated native tool built upon the same libraries.

The measured aperture flux densities agree closely, typically differing by only a few percent. Both systems report local background estimates, background RMS, peak signal-to-noise ratios, and source-size information alongside the measured flux densities and uncertainties. \maasai{} additionally provides structured provenance metadata describing the synthesised beam, pixel scale, background-estimation procedure, and measurement reliability indicators, together with downloadable native-tool artefacts.

The reported morphology labels differ because the systems employ different strategies: GPT-5.5 derives them from geometric heuristics based on the measured aperture size relative to the synthesised beam and the peak signal-to-noise ratio, whereas \maasai{} preserves the annotations explicitly provided in the input DS9 region metadata, reporting their provenance through a dedicated field.

\paragraph{Source cutout archive retrieval}\leavevmode%
\label{subsec:source-cutout-task}
We evaluated archive-cutout retrieval by requesting FITS cutouts from selected astronomical surveys (WISE W3 band, Hi-GAL 70~$\mu$m, 2MASS J band, and SDSS g band) for each source region identified in a radio image. Cutout dimensions were determined by the region extent, and a detailed retrieval report was required for each survey. For this test, we used the radio source regions shown in Fig.~\ref{fig:sample-img-regions}. The complete task prompt is reported in \ref{app:source-cutout-prompt}.

Both \maasai{} and GPT-5.5 correctly interpreted the supplied DS9 regions, derived the source positions and angular extents, and generated structured reports for every requested region--survey combination. GPT-5.5 dynamically generated Python code to query public archives, but execution-environment limitations prevented retrieval of the WISE and 2MASS cutouts, which were reported as unsuccessful. In contrast, \maasai{} delegated the task to a validated native archive-retrieval tool, successfully downloading the WISE and 2MASS cutouts while correctly reporting the lack of SDSS coverage and the current absence of Hi-GAL support. Both systems preserved the requested survey list without substituting alternative datasets. \maasai{} additionally produced a downloadable archive of the retrieved FITS products together with a complete JSON manifest describing the provenance and status of every retrieval.

\subsection{Multi-step analysis tasks}
\label{subsec:multi-step-analysis-tasks}
We evaluate multi-step scientific requests requiring the orchestration of dependent operations and the exchange of intermediate data products, including a quantitative image-analysis benchmark and representative workflows involving source extraction, archive queries, and catalogue comparison.

\subsubsection{B3 - Joint image classification and source detection}
\label{subsec:benchmark_B3}
Benchmark B3 evaluates a composite radio-image analysis task combining image-level classification, source detection, and morphological characterisation. Unlike benchmarks~B1 and~B2, this task requires three distinct outputs for each input image: identification of visible imaging artefacts, assignment of a global peculiarity class among \texttt{ORDINARY}, \texttt{COMPLEX}, and \texttt{PECULIAR}, and detection and classification of statistically significant source-like objects. Each detected object must be localised using an absolute-pixel bounding box and assigned one of the morphological classes \texttt{COMPACT}, \texttt{EXTENDED}, \texttt{EXTENDED-MULTISLAND}, \texttt{SPURIOUS}, \texttt{FLAGGED}, or \texttt{UNKNOWN}\footnote{\texttt{SPURIOUS} denotes false detections generated by imaging artefacts, while \texttt{FLAGGED} denotes real sources whose morphology is substantially contaminated by nearby artefacts.}.

For this task, we used 450 images from the ASKAP EMU Pilot Survey \citep{Norris2021}, selected from the \emph{radioimg-multilabel} test dataset described in \cite{Riggi2025}. These images were additionally annotated in this work with source bounding boxes and morphological classes, providing ground truth for all three subtasks.

The complete task prompt, including class definitions and the output schema, is reported in \ref{app:benchmark-b3-prompt}.

\begin{table}[!ht]
\centering
\caption{Performance comparison between GPT-5.5 and \maasai{} on the B3 multi-step radio-image analysis benchmark. Both systems are evaluated on the same set of 450 images. Source-detection metrics are class-agnostic at
an IoU threshold of 0.5, while source-classification metrics are computed over geometrically matched detections. Best values are shown in bold.}
\label{tab:b3_results}
\footnotesize
\begin{tabular}{llcc}
\hline
\textbf{Task} & \textbf{Metric} & \textbf{GPT-5.5} (\%) & \textbf{\maasai{}} (\%)\\
\hline

Artefact classification
	& Accuracy & 83.6 & \textbf{85.6} \\
	& Macro F1 & 83.0 & \textbf{84.4} \\

\hline

Peculiarity classification
	& Accuracy & 67.8 & \textbf{76.2} \\
	& Macro F1 & 69.2 & \textbf{77.4} \\

\hline

Source detection
	& Precision & 79.1 & \textbf{88.0} \\
	& Recall & 49.2 & \textbf{58.4} \\
	& F1 & 60.7 & \textbf{70.2} \\
	& Mean IoU & 71.6 & \textbf{75.0} \\

\hline

Source classification
	& Accuracy & 94.2 & \textbf{97.1} \\
	& Macro F1 & 55.3 & \textbf{60.8} \\

\hline
\end{tabular}
\end{table}

Table~\ref{tab:b3_results} reports the achieved results. 
\maasai{} outperforms GPT-5.5 across all three subtasks, with improvements in macro F1 of 1.4 and 8.2 percentage points for artefact and image peculiarity classification, respectively. The class-wise peculiarity results reveal different performance patterns: GPT-5.5 more reliably identifies \texttt{ORDINARY} images, whereas \maasai{} achieves higher recall for \texttt{COMPLEX} (94\% vs 75\%) and \texttt{PECULIAR} (60\% vs 48\%) images. Nevertheless, confusion between these two classes remains the main limitation for \maasai{}, with approximately 38\% of \texttt{PECULIAR} images misclassified as \texttt{COMPLEX}.

The largest improvement in F1 score is observed for source detection, with an increase of 9.5 percentage points, accompanied by gains in both precision and recall. The mean IoU of matched detections also increases by 3.4 percentage points, indicating improved source localisation. However, the detection gains are morphology dependent: \maasai{} achieves higher class-agnostic detection recall for the dominant \texttt{COMPACT} population (62\% vs 52\%), whereas GPT-5.5 performs better on \texttt{EXTENDED} sources (84\% vs 73\%). For geometrically matched sources, \maasai{} also improves morphological classification accuracy (97.1\% vs 94.2\%) and macro F1 (60.8\% vs 55.3\%).

Despite the high overall accuracy for matched sources, classification of less populated morphological classes remains challenging, particularly \texttt{EXTENDED-MULTISLAND}.
Among the 48 geometrically matched sources of this class, \maasai{} misclassifies 35 (73\%) as \texttt{EXTENDED} and correctly identifies 12 (25\%), with the remaining source assigned to another class. A similar confusion is observed for GPT-5.5, suggesting that distinguishing multi-island sources from ordinary extended morphologies remains difficult for both systems.

In this benchmark, \maasai{} performs source detection through a specialised radio-source extraction tool, whereas GPT-5.5's internal procedure in Batch API mode is not exposed. It is therefore unclear whether GPT-5.5 performs source localisation directly or through generated code invoking conventional source-finding algorithms, as observed in related experiments using the ChatGPT web interface (Section~\ref{subsec:multi-step-native-tasks}). The results should consequently be interpreted as an end-to-end system comparison rather than a comparison of specialised tool execution with unaided visual reasoning.

\begin{table*}[!ht]
\centering
\caption{Summary of representative multi-step native-tool demonstrations. These experiments assess the orchestration of multiple astronomical operations and are interpreted as capability demonstrations rather than quantitative benchmarks.}
\label{tab:multi_step_native_tools}
\footnotesize%
\begin{tabular}{p{2.5cm} p{4.4cm} p{4.3cm} p{5.5cm}}
\hline
\textbf{Scientific task} &
\textbf{Orchestrated capabilities} &
\textbf{Typical output} &
\textbf{Comparison with GPT-5.5} \\
\hline

Archive-assisted source identification &
Radio-source extraction followed by positional cross-matching with an astronomical object archive &
Detected-source catalogue with image and sky coordinates, candidate identifiers and object types, angular separations, and annotated JSON/CSV cross-match products &
Successfully extracted sources and generated structured catalogues, but could not complete the SIMBAD queries because the service endpoint was inaccessible; \maasai{} completed the end-to-end workflow and returned positional counterpart candidates.\\

\hline%

Catalogue comparison &
Independent source extraction with two finders, one-to-one positional association, statistical comparison, and merged-catalogue generation &
Common and finder-specific detections, positional offsets, matching fractions, summary metrics, and a merged catalogue preserving measurements and provenance &
Generated a complete comparison report and merged catalogue through dynamically generated and adapted Python code. \maasai{} instead executed an explicit three-step workflow using \texttt{caesar}, \texttt{Aegean}, and a native catalogue-comparison tool, preserving the intermediate products and provenance of each stage.\\

\hline
\end{tabular}
\end{table*}

\subsubsection{Additional multi-step tool demonstrations}
\label{subsec:multi-step-native-tasks}
We additionally evaluated two representative workflows combining source extraction with archive cross-matching or catalogue comparison. These demonstrations assess the orchestration of specialised tools and the preservation of intermediate scientific products. Table~\ref{tab:multi_step_native_tools} summarises the workflows and their qualitative comparison with GPT-5.5.

\begin{figure}[!htb]
\centering%
\includegraphics[scale=0.38]{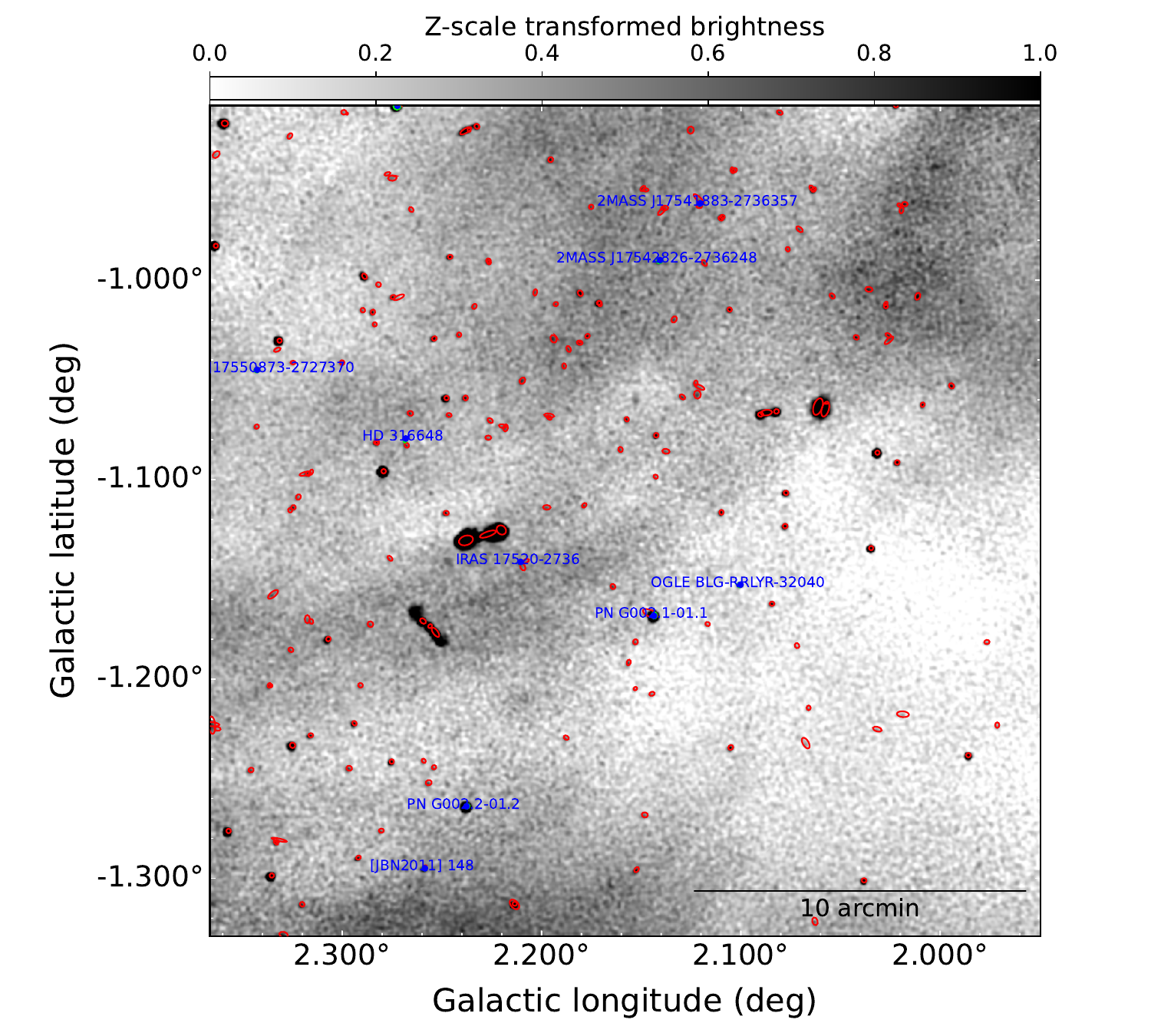}
\caption{Sample radio image, taken from the MeerKAT SMGPS survey \citep{SMGPS}, with superimposed sources extracted and cross-matched against the SIMBAD archive using a 2-arcsec matching radius. Catalog results were obtained by \maasai{} through astronomical source extraction (\texttt{caesar}) and archive query tools. Sources shown in red have no matching SIMBAD entry, whereas sources shown in green have at least one SIMBAD counterpart, shown in blue.}%
\label{fig:sfind-crossmatch-maasai-results}
\end{figure}

\paragraph{Archive-assisted source identification}\leavevmode%
\label{subsec:data-retrieval-task}
We evaluated the orchestration of source extraction and archive cross-matching using a request that required a conventional radio-source finder, a fixed 2-arcsec SIMBAD matching radius, and structured reporting of both candidate and unmatched sources. The complete task prompt is reported in \ref{app:archive-source-identification-prompt}.

Both \maasai{} and GPT-5.5 successfully performed the source-extraction stage using established radio-astronomy software. However, only \maasai{} completed the subsequent SIMBAD cross-matching stage. \maasai{} created a two-step plan using the \texttt{caesar} source finder followed by the native archive cross-matching tool. A total of 201 sources were detected, with positional counterpart candidates found in SIMBAD for 10 of them. Results are shown in Fig.~\ref{fig:sfind-crossmatch-maasai-results}.

GPT-5.5, on the other hand, dynamically generated Python code employing the \texttt{Aegean} source finder for the extraction stage. A total of 138 catalogue components were detected. In all performed runs, cross-matching information was explicitly reported as as unavailable because the SIMBAD programmatic query service was inaccessible from its execution environment.

Overall, the different source counts reflect the use of distinct extraction and component-fitting methods, with \maasai{} providing the more complete end-to-end result by returning counterpart identifiers, object types, separations, annotated catalogues, and downloadable cross-match tables.

\paragraph{Catalog comparison}\leavevmode%
\label{subsec:catalog-analysis-task}
We evaluated the orchestration of independent source extraction and catalogue comparison using a request that required two conventional source finders, one-to-one positional matching within a fixed 4-arcsec radius, summary statistics, and a merged catalogue preserving the measurements and provenance of both inputs. The complete task prompt is reported in \ref{app:catalog-comparison-prompt}.

\maasai{} executed a three-step workflow, sequentially invoking \texttt{caesar}, \texttt{Aegean}, and a native catalogue-comparison tool to generate matched and merged catalogues together with summary statistics. GPT-5.5 instead synthesised Python code that automatically discovered and employed available astronomical software (including \texttt{PyBDSF} and \texttt{Aegean}). 
When the full \texttt{Aegean} fitting stage exceeded the available execution time, it switched to an alternative implementation based on Gaussian fitting and seeded island segmentation, explicitly documenting this choice. Both approaches produced complete comparison reports and machine-readable output catalogues.

\begin{figure*}[!htb]
\centering%
\includegraphics[width=\textwidth]{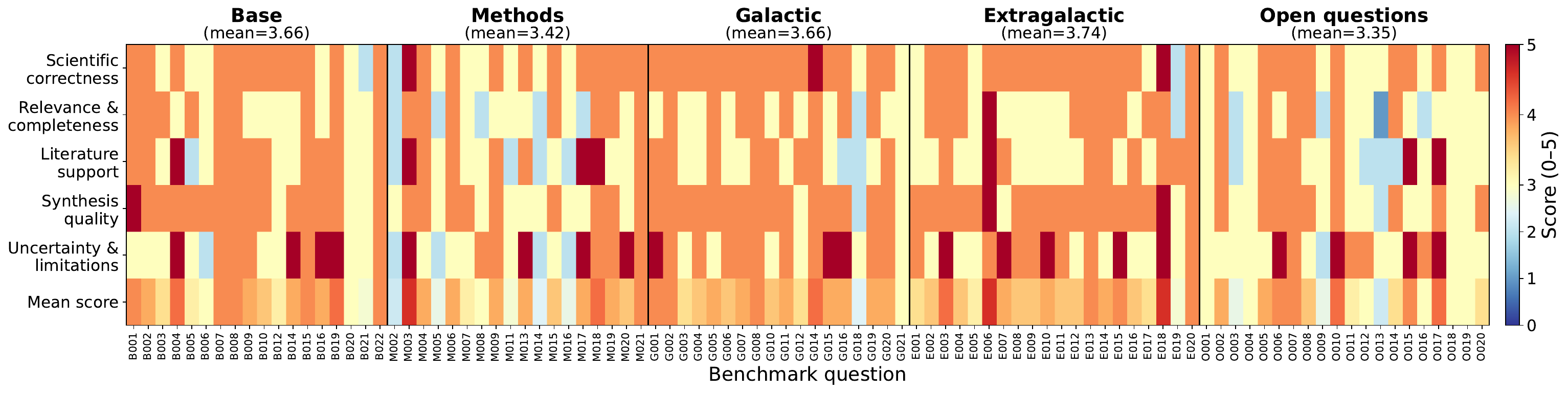}
\caption{Literature benchmark results: judge scores by question and criterion. Columns represent individual benchmark questions, grouped into five thematic areas, while rows show the five evaluation criteria and the mean score across criteria. The mean score for each thematic area is reported above the corresponding group. Scores range from 0 to 5, with higher values indicating better performance.}%
\label{fig:literature-judge-results}
\end{figure*}

\begin{figure*}[!htb]
\centering%
\includegraphics[width=\textwidth]{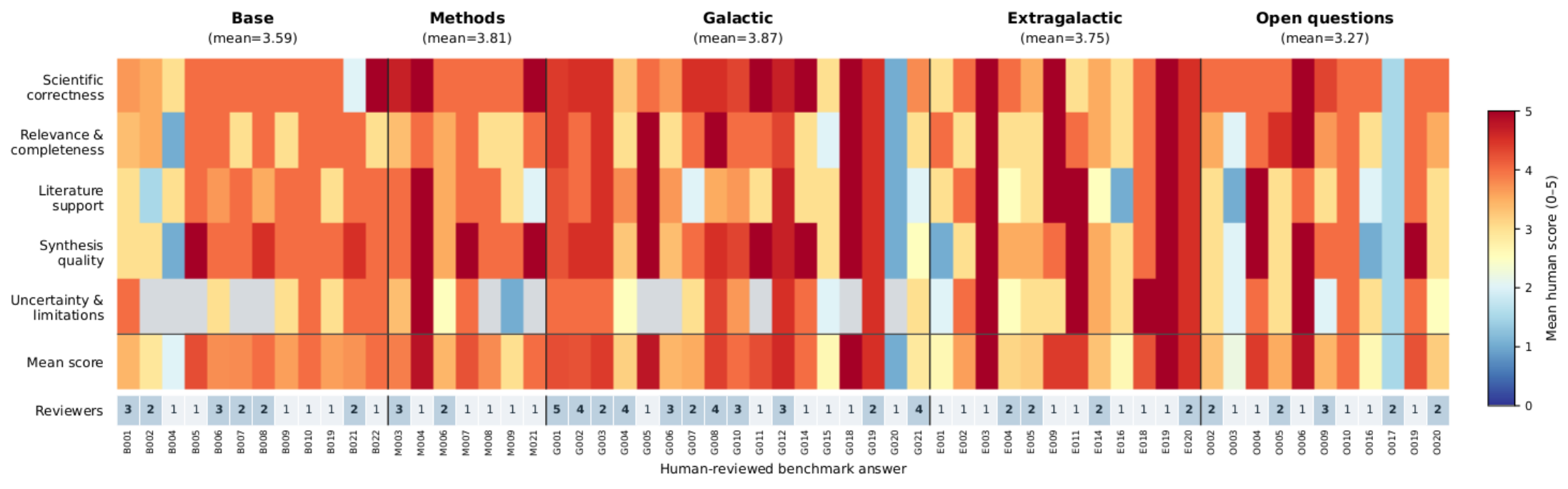}
\caption{Human evaluation scores for the subset of 59 benchmark answers reviewed by astronomers, grouped by thematic area. Each cell represents the mean score assigned by the available human reviewers to a given answer and evaluation criterion, on a scale from 0 to 5. The bottom heatmap row shows the mean across available criteria, while the additional row indicates the number of independent reviewers for each answer. Grey cells denote missing or not-applicable ratings. The displayed means are descriptive summaries and do not represent inter-rater agreement.}%
\label{fig:literature-human-results}
\end{figure*}

\subsection{Literature analysis tasks}
\label{subsec:literature-analysis-tasks}
To assess \maasai{}'s ability to answer scientific questions through literature retrieval and synthesis, we constructed a benchmark covering five thematic areas: basic radio-astronomy knowledge; observational methods and analysis techniques; Galactic radio astronomy; extragalactic radio astronomy; and open questions in radio astronomy.

An initial pool of 100 questions (20 per thematic area) was generated using two deep-research models, OpenAI GPT-5.5 Pro and Google Gemini 3.5 Thinking, and subsequently reviewed by astronomers for scientific relevance, clarity, and suitability for literature-based analysis. Questions were revised, removed, or replaced where necessary to reduce duplication and improve thematic coverage, with additional questions introduced by domain experts. The final benchmark comprises 94 questions. Representative examples are reported in~\ref{app:literature-questions}, and the complete benchmark is available in the project software repository.

Each question was submitted independently to \maasai{} using the prompt reported in~\ref{app:literature-analysis-prompt}. The system was instructed to produce a concise scientific synthesis grounded in evidence retrieved from its literature knowledge base (Section~\ref{subsec:rag}), integrating relevant findings, distinguishing established results from interpretations or unresolved aspects, and acknowledging insufficient evidence. Citations were restricted to the retrieved sources.

The resulting answers were evaluated by GPT-5.5 acting as an automated LLM judge, using the evaluation prompt reported in~\ref{app:literature-evaluation-prompt}. Five criteria were scored independently on an integer scale from 0 to 5: (i) \emph{scientific correctness}, assessing the accuracy of scientific statements and conclusions; (ii) \emph{relevance and completeness}, assessing coverage of the question's important aspects; (iii) \emph{literature support}, assessing the evidence supporting substantive claims and their citations; (iv) \emph{synthesis quality}, assessing the coherent integration of retrieved evidence; and (v) \emph{uncertainty and limitations}, assessing the appropriate treatment of assumptions, disagreements, and insufficient evidence. A subset of the answers was also independently evaluated by astronomers using the same criteria and supporting literature, enabling an exploratory comparison between automated and expert assessments.

\subsubsection{Model evaluation}
\label{subsec:literature-analysis-tasks-model-eval}
The model-based evaluation results are summarised in Fig.~\ref{fig:literature-judge-results}. Across the 94 benchmark questions, the mean criterion-averaged score was 3.56 out of 5. Mean scores were 3.69 for scientific correctness, 3.32 for relevance and completeness, 3.42 for literature support, 3.71 for synthesis quality, and 3.68 for uncertainty and limitations. 
Synthesis quality received the highest mean score, followed closely by scientific correctness and uncertainty and limitations, with 70\% of responses receiving a synthesis-quality score of at least 4. Uncertainty and limitations also received a comparatively high mean score, although with the largest question-to-question dispersion.

Relevance and completeness, together with literature support, were the comparatively weaker dimensions, with approximately 11\% of answers receiving scores of 2 or below for each criterion. These results may partly reflect the constraints imposed on response generation, including a maximum length of 350 words and a limit of eight retrieved literature references. Introduced to facilitate a manageable human evaluation, these restrictions may have limited the coverage of questions requiring more extensive explanations or literature synthesis, as also suggested by feedback from human reviewers. Future evaluations could therefore explore question-dependent limits informed by scientific complexity. Overall, the automated assessment indicates generally favourable ratings for the coherence and scientific content of the generated answers, while identifying completeness and literature support as the principal areas for improvement.

\subsubsection{Human evaluation}
\label{subsec:literature-analysis-tasks-human-eval}
We conducted a pilot evaluation in which astronomers with expertise in different areas of radio astronomy independently reviewed the generated answers through a dedicated web interface. At the time of this analysis, 59 answers had received at least one evaluation, including 28 evaluated by two or more reviewers, for a total of 105 individual reviews.

\paragraph{Human assessments and inter-reviewer reliability}
The human evaluation scores are summarised in Fig.~\ref{fig:literature-human-results}. The mean criterion-averaged score across the reviewed answers was 3.69 out of 5, averaging the available ratings within each answer before computing the overall mean. Mean scores for scientific correctness, relevance and completeness, literature support, synthesis quality, and uncertainty and limitations were 4.00, 3.65, 3.49, 3.82, and 3.49, respectively. Human assessments were therefore generally positive, particularly for scientific correctness and synthesis quality, although these averages do not capture variability between independent reviewers.

We first examined the distribution of absolute score differences between independent human reviewers. As shown in Fig.~\ref{fig:literature-human-pair-differences}, exact agreement (absolute difference of 0) ranged from 18.3\% to 32.4\% across the five evaluation criteria, while the proportion of reviewer pairs differing by at most one score point (absolute difference of 0 or 1) ranged from 64.8\% to 76.1\%. Thus, although identical ratings were relatively infrequent, most reviewer pairs assigned scores that differed by no more than one point, indicating fairly common near-agreement.

To assess the reproducibility of human ratings beyond chance expectations, we computed Krippendorff's ordinal alpha ($\alpha_{\mathrm{ordinal}}$). Negative values indicate greater disagreement than expected by chance, zero corresponds to chance-level agreement, and unity represents perfect agreement (see~\ref{app:literature-evaluation-statistics} for details on the statistical measures used throughout this subsection). The analysis included 28 answers (74 ratings) for the first four criteria and 20 answers (50 ratings) for uncertainty and limitations, after excluding not-applicable ratings. Agreement estimates remained low, with $\alpha_{\mathrm{ordinal}}$ ranging from $-0.025$ for scientific correctness to $0.313$ for uncertainty and limitations (Fig.~\ref{fig:literature-human-alpha}). The 95\% question-level bootstrap confidence intervals encompassed zero for four criteria, while the interval for uncertainty and limitations had a lower bound marginally above zero. 

Overall, these results indicate limited and uncertain inter-reviewer reliability, despite relatively frequent agreement within one score point and a modest positive agreement estimate for uncertainty and limitations.

\paragraph{Human--LLM comparison}
We next compared individual human ratings with the corresponding automated judge scores, obtaining 105 human--LLM rating pairs across 59 answers for the first four criteria and 77 pairs across 47 answers for uncertainty and limitations.

The paired-score distributions are shown in Fig.~\ref{fig:literature-human-llm-score-map}, with the diagonal indicating exact agreement and the off-diagonal cells showing the magnitude and direction of scoring differences. Exact agreement ranged from 26.7\% to 39.0\% across the five evaluation criteria, increasing to 73.3--88.6\% for scores differing by at most one point. Thus, human and LLM ratings were frequently close, although identical scores were considerably less common.

We then examined potential systematic scoring differences by subtracting the mean human rating from the LLM score for each answer and averaging these differences across answers, so that each answer contributed once regardless of the number of reviewers. The largest negative mean differences were observed for scientific correctness ($-0.233$) and relevance and completeness ($-0.289$), suggesting slightly lower LLM scores on average for these criteria. Differences for literature support ($+0.074$), synthesis quality ($+0.012$), and uncertainty and limitations ($+0.152$) were smaller. The 95\% question-level bootstrap confidence interval excluded zero only for relevance and completeness ($[-0.538,-0.036]$; Fig.~\ref{fig:literature-human-llm-bias}). Exploratory analyses also suggested variation across thematic areas (Fig.~\ref{fig:literature-human-llm-bias-thematic}), although the limited samples and multiple comparisons preclude robust domain-specific conclusions.

To assess agreement beyond chance expectations, we computed quadratic weighted Cohen's kappa ($\kappa_w$), which penalises larger score differences more strongly. As for Krippendorff's alpha, zero indicates chance-level agreement, unity represents perfect agreement, and negative values indicate greater disagreement than expected by chance. The resulting coefficients were low ($-0.036$ to $0.141$), indicating limited chance-corrected agreement despite the relatively frequent differences of at most one score point. These pooled agreement statistics are descriptive, since the human ratings originate from different reviewers and multiple rating pairs may correspond to the same answer.

We also examined whether the LLM and human reviewers ranked the answers similarly, comparing the LLM score with the median human rating for each answer. Spearman's rank correlations were weak across all five evaluation criteria ($\rho=-0.135$ to $0.197$), with Kendall's $\tau_b$ yielding consistent results. This indicates that answers receiving higher LLM scores did not consistently receive higher scores from human reviewers, and vice versa.

\paragraph{Summary of findings}
The main findings of this pilot evaluation can be summarised as follows:

\begin{itemize}
\item \textit{Human assessment}: Astronomers generally assigned favourable scores to the generated answers, particularly for scientific correctness and synthesis quality.

\item \textit{Inter-reviewer reliability}: Human reviewers frequently assigned scores differing by at most one point, but chance-corrected agreement estimates were generally low and uncertain, indicating limited evidence of reproducible scoring across reviewers.

\item \textit{Human--LLM agreement}: Automated and human scores frequently differed by at most one point, but chance-corrected agreement and rank associations were weak. The automated judge tended to assign slightly lower mean scores for scientific correctness and relevance and completeness, although only the latter showed a bootstrap confidence interval excluding zero. Differences for the remaining criteria were relatively small.
\end{itemize}

Given the limited inter-reviewer reliability and the preliminary nature of the human evaluation, these findings do not establish whether the automated judge is less consistent than human experts. Additional independent human assessments will be necessary to obtain more robust estimates of agreement between automated and expert evaluations.


\section{Conclusions and outlook}
\label{sec:conclusions}
We have presented \maasai{}, a modular multi-agent system for astronomical data analysis that combines foundation-model reasoning, retrieval-augmented scientific knowledge, and specialised astronomical tools within a graph-based execution framework. The system coordinates request preparation, planning, execution, and response synthesis through specialised workers, while supporting heterogeneous scientific data products, explicit provenance management, and configurable inference backends.

We evaluated \maasai{} through quantitative radio-image benchmarks and representative scientific workflows. Compared with GPT-5.5, \maasai{} achieved higher macro F1 scores for extended/diffuse source classification (0.72 vs 0.53) and source morphology classification (0.88 vs 0.37), while source-detection F1 in the composite benchmark increased from 60.7\% to 70.2\%. Performance nevertheless varied with source morphology, with extended and multi-island sources remaining challenging in some tasks. These end-to-end comparisons do not isolate the contributions of individual system components. Complementary demonstrations showed the orchestration of heterogeneous astronomical tools and archives across single- and multi-step workflows, with intermediate products and provenance preserved.

Literature-grounded reasoning was evaluated on 94 radio-astronomy questions, with generally favourable automated assessments and a pilot human evaluation covering 59 answers. Scientific correctness and synthesis quality received relatively high ratings, while completeness and literature support remained weaker dimensions. However, limited inter-reviewer reliability and weak agreement between automated and human scores highlight the need for further expert evaluation.

\smallskip

Several limitations and directions for further development have been identified:

\begin{itemize}
\item \textit{Computational efficiency}: Multi-stage workflows require multiple sequential model inferences, introducing computational latency that can reach several minutes with the locally deployed model used in this work. The separation between orchestration logic and inference backends allows newer models and serving frameworks to be incorporated without redesigning the workflow. 
Advances in inference hardware, model efficiency, and serving infrastructure are therefore expected to help reduce execution times.

\item \textit{Cross-domain extensions}: The present evaluation has focused primarily on radio astronomy, while the system's capabilities across other scientific domains remain to be systematically assessed. However, the implementation is not intrinsically tied to radio astronomy, and some initial cross-domain and multimodal capabilities are already available, including literature retrieval for solar physics and exoplanet science, solar-flare forecasting, and tabular and light-curve data processing. Extending these capabilities into complete and systematically evaluated workflows therefore represents a natural next step.

\item \textit{Literature retrieval and evaluation}: The literature-analysis results motivate further investigation of retrieval strategies and their dependence on scientific corpora and query characteristics, together with more extensive expert evaluation of the generated responses.

\item \textit{Advanced scientific workflow orchestration}:
Ongoing developments focus on supporting more complex, multi-stage scientific pipelines through intermediate result evaluation, adaptive execution decisions, persistent workflow state, and human-in-the-loop feedback. These capabilities are being integrated with the \texttt{caesar-rest} pipeline infrastructure, with radio-interferometric calibration and imaging among the primary target applications.

\item \textit{Deployment and scalability}:
The system has so far been deployed and evaluated on relatively small-scale computing infrastructure and with a limited user base, while its operation within larger, distributed research infrastructures remains to be systematically assessed. Ongoing developments aim to support deployment within larger INAF computing infrastructures, providing a basis for investigating the integration of agent-assisted scientific workflows within environments such as the SKA Regional Centre Network (SRCNet).
\end{itemize}

\medskip%

This work illustrates how the complementary capabilities of foundation models and specialised scientific software can be combined within a unified agentic framework, allowing astronomical analyses to be initiated and coordinated through natural-language interaction while delegating domain-specific computations to established scientific tools. The separation between orchestration logic, scientific capabilities, and inference backends facilitates the integration of new models and astronomical software, while also supporting fully local deployment in environments with data-governance or institutional requirements. As the volume and heterogeneity of astronomical data continue to grow, particularly with the advent of the SKA, the ability to compose, inspect, and extend such tool-augmented workflows may play an important role in exploiting the scientific capabilities of next-generation facilities.

\section*{Acknowledgements}
\label{sec:acknowledgements}
\small{
This work was carried out as part of the MAASAI (\emph{Multi-agent Astronomy Assistant for Scientific Analysis and Inference}) research initiative. We acknowledge the INAF projects "CIRASA" (\emph{Collaborative and Integrated platform for Radio Astronomical Source Analysis}) and "SCIARADA" (\emph{Self-supervised Contrastive learning for Inspection and Analysis of RAdio DAta in the SKA era}), the Italian PNRR project IR0000034 "STILES" (\emph{Strengthening the Italian leadership in ELT and SKA}), the Italian PON 2014--2020 project "MOSAICO", and the \textit{Laboratory for Interdisciplinary Research in Astronomy and AI} (LIRA) at INAF-OACT for providing the computational resources used in this work.

S.R., S.E.M. and T.S acknowledge the INAF "WildKAT" (\emph{Wide-field Intelligent Learning \& Discovery with MeerKAT and precursors}) project for financial support. 

The authors acknowledge the developers of the software tools and libraries extensively used in this work, including \texttt{LangChain}/\texttt{LangGraph} (agent system development), \texttt{astropy}~\citep{astropy2013,astropy2018}/\texttt{astroquery}~\citep{astroquery} (astronomical analysis tools), and \texttt{OpenAI ChatGPT}/\texttt{Google Gemini} (assistance with manuscript language editing, code development, and testing).
}%

\section*{Data Availability}
\label{sec:data-availability}
\small{
The software developed in this work will be made publicly available under the GNU General Public License v3.0\footnote{\scriptsize{\url{https://www.gnu.org/licenses/gpl-3.0.html}}} upon acceptance of the manuscript for publication, at \url{https://github.com/inaf-oact-ai/maasai/}.
}%







\newpage%

\appendix%
\setcounter{figure}{0}
\setcounter{table}{0}

\onecolumn%

\section{Prompts for single-step radio data analysis}
\label{app:single-step-prompts}
This section reports the complete prompts used for the single-step quantitative benchmarks and native-tool capability demonstrations discussed in Section~\ref{subsec:single-step-analysis-tasks}.

\subsection{B1 - Extended/diffuse source detection}
\label{app:benchmark-b1-prompt}
The following prompt was used for the extended/diffuse source detection task (benchmark B1):
\begin{texthighlight}
### Context: Consider these morphological classes of radio astronomical sources, defined as follows: 
 EXTENDED: This class comprises either single-island compact objects with sharp edges, having a morphology and size dissimilar to that of the image synthesised beam (e.g. 10 times larger than the beam size or with elongated shape), or disjoint multi-island objects, where each island can have either a compact or extended morphology and can host single or multiple emission components. Typical examples are extended radio galaxies formed by a single elongated island or by multiple islands, hosting the galaxy core and lobe structures;
 DIFFUSE: a particular class of single-island extended objects with small angular size (e.g. smaller than few arcminutes), having diffuse edges and a roundish morphology; 
 DIFFUSE-LARGE: large-scale (e.g. larger than few arcminutes and covering a large portion of the image) diffuse object with irregular shape.

An island is a group of 4-connected pixels in an image under analysis with intensity above a detection threshold with respect to the sky background level.

### Question: Which of these morphological classes of radio sources do you see in the image?
EXTENDED
DIFFUSE
DIFFUSE-LARGE

Answer the question using the provided context. Report the identified class labels separated by commas, without any additional explanation text. Report just NONE if you cannot recognise any of the above classes in the image.
\end{texthighlight}

\subsection{B2 - Source morphology classification}
\label{app:benchmark-b2-prompt}
The following prompt was used for the radio source morphology classification task (benchmark B2):
\begin{texthighlight}
### Context: Consider these morphological classes of radio astronomical sources:
 1C-1P: single-island sources having only one flux intensity peak;
 1C-2P: single-island sources having two flux intensity peaks; 
 1C-3P: single-island sources having three flux intensity peaks;
 2C-2P: sources consisting of two separated islands, each hosting a single flux intensity peak;
 2C-3P: sources consisting of two separated islands, one containing a single peak of flux intensity and the other exhibiting two distinct intensity peaks;
 3C-3P: sources consisting of three separated islands, each hosting a single flux intensity peak. 

An island is a group of 4-connected pixels in an image under analysis with intensity above a detection threshold with respect to the sky background level.

### Question: Which of these morphological classes of radio sources do you see in the image?
1C-1P
1C-2P
1C-3P
2C-2P
2C-3P
3C-3P

Answer the question using the provided context. Report only the identified class label, without any additional explanation text.
\end{texthighlight}

\subsection{Source measurements}
\label{app:source-measurement-prompt}
The following prompt was used for the aperture-photometry and source-measurement demonstration:
\begin{texthighlight}
Compute the aperture photometry of the radio sources defined in the attached DS9 region file on the attached FITS image. Estimate the local background for each source, compute the background-subtracted flux density and its uncertainty, classify the source morphology, and return a summary table with the measured quantities.
\end{texthighlight}

\subsection{Source cutout archive retrieval}
\label{app:source-cutout-prompt}
The following prompt was used for the multi-survey archival-cutout retrieval demonstration:
\begin{texthighlight}
Retrieve archival FITS image cutouts around every source defined in the attached DS9 region file using WISE W3, Hi-GAL 70 micron, 2MASS J, and SDSS g surveys.

For each region, use the attached FITS image as the WCS reference if necessary, determine the sky center and angular extent, and request a square cutout whose total width is 20\% larger than the maximum source extent.

Report the result for every requested region--survey combination, including missing coverage and failed retrievals. Do not replace unavailable surveys with alternatives.

Return the region identifier, coordinates, original extent, requested size, survey, wavelength domain, retrieval status, and output filename. Save a complete JSON manifest and bundle all successfully retrieved FITS products into a downloadable archive.
\end{texthighlight}

\section{Prompts for multi-step radio data analysis}
\label{app:multi-step-prompts}
This section reports the complete prompts used for the quantitative multi-stage benchmark and the native-tool workflow demonstrations discussed in Section~\ref{subsec:multi-step-analysis-tasks}.

\subsection{Joint image classification and source detection}
\label{app:benchmark-b3-prompt}
The following prompt was used for the joint artefact assessment, image-level peculiarity classification, and source detection task (benchmark B3).

\begin{texthighlight}
### GLOSSARY

- Synthesized beam: the effective angular resolution and directional response pattern of a radio telescope array operating via aperture synthesis. It acts as the instrument point-spread function, defining how an unresolved point source appears in the radio image.
- Island: a group of 4-connected pixels whose intensity is above a detection threshold with respect to the local sky background.
- Imaging artefacts: visible non-astrophysical and noise patterns caused by the imaging or deconvolution process. They may appear as rings, stripes, sidelobes, fragmented structures, or elongated or repeated patterns, often around bright sources. They are often mistaken for real radio sources but are actually spurious.

### SOURCE MORPHOLOGY CLASSES

Consider the following possible morphological classes of radio sources:

- COMPACT: an isolated radio source consisting of a single island with a point-like or slightly resolved morphology. The source may contain one or more blended emission components whose shapes are broadly consistent with the synthesized beam.
- EXTENDED: a radio source consisting of a single island with resolved extended morphology. The source may contain one or more blended components, at least some of which deviate significantly from the synthesized beam shape.
- EXTENDED-MULTISLAND: a radio source composed of multiple disconnected islands of emission. Each island may contain one or more compact or extended components.
- SPURIOUS: a false source-like detection associated with imaging artefacts rather than a real astrophysical source.
- FLAGGED: a real single-island radio source, with compact or extended morphology, that is poorly imaged and significantly contaminated by nearby overlapping imaging artefacts.
- UNKNOWN: a detected source whose morphology cannot be determined reliably.

### IMAGE-LEVEL PECULIARITY CLASSES

A radio image can be assigned one global peculiarity class:

- ORDINARY: the image contains only background, imaging-artefact patterns, or point-like or slightly resolved compact radio sources, without clearly extended or unusual astrophysical emission.
- COMPLEX: the image contains one or more extended, multi-island, diffuse, or radio-galaxy-like sources, but no strongly anomalous or peculiar morphology.
- PECULIAR: the image contains one or more anomalous, unusual, or strongly peculiar radio structures, such as very diffuse emission, irregular extended shapes, filamentary, ring-like, or arc-like structures, or emission covering a large portion of the image.

### TASK

Analyse the input radio image. The task has three distinct stages and corresponding outputs, which may require different analysis operations:

1. Decide whether the image contains visible imaging artefacts. Return YES if artefacts are visible, otherwise NO.
2. Assign one global peculiarity class to the image: ORDINARY, COMPLEX, or PECULIAR.
3. Identify all source-like objects whose emission is significantly above the local sky background, for example at least $5\sigma$, and produce a source catalogue. For each catalogued object, estimate its bounding box and assign the most appropriate source morphological class.

### OUTPUT FORMAT

Return only a valid JSON object following exactly this schema:

{
  "has_artefacts": "NO",
  "peculiarity_class": "ORDINARY",
  "sources": [
    {
      "class_label": "COMPACT",
      "x1": 10.0,
      "x2": 30.0,
      "y1": 40.0,
      "y2": 60.0
    }
  ]
}

where:

- has_artefacts must be exactly one of: YES, NO.
- peculiarity_class must be exactly one of: ORDINARY, COMPLEX, PECULIAR.
- sources must be a list of detected source-like objects.
- class_label must be exactly one of: UNKNOWN, SPURIOUS, COMPACT, EXTENDED, EXTENDED-MULTISLAND, FLAGGED.
- Use SPURIOUS for source-like false detections caused by artefacts.
- Use FLAGGED for real sources heavily contaminated by nearby artefacts.
- Use UNKNOWN if a detected object's morphology cannot be determined reliably.
- x1 and x2 are the minimum and maximum pixel coordinates along the horizontal image axis.
- y1 and y2 are the minimum and maximum pixel coordinates along the vertical image axis.
- Bounding-box coordinates must be absolute image pixel coordinates and satisfy x1 < x2 and y1 < y2.
- If no source-like objects are detected, return an empty list: "sources": [].

Return only the JSON object and no additional text.
\end{texthighlight}

\subsection{Archive-assisted source identification}
\label{app:archive-source-identification-prompt}

The following prompt was used for the source-extraction and SIMBAD positional-cross-matching demonstration:

\begin{texthighlight}
Detect the radio sources in the attached FITS image and search for known astronomical objects near their sky positions.

Use a conventional radio-source extraction and fitting tool that produces a source catalogue containing both image and sky coordinates.

Adopt a fixed cross-match radius equal to 2 arcsec and use it consistently for all detected sources.

Cross-match the resulting source catalogue against SIMBAD archive. For each detected source:
- report its source identifier;
- report its image coordinates when available;
- report its right ascension and declination;
- report the nearest SIMBAD candidate within the adopted search radius;
- report the candidate identifier, object type, and angular separation;
- report any additional candidates when more than one object is returned;
- explicitly indicate when no SIMBAD candidate is found within the search radius.

Treat the returned SIMBAD objects as positional counterpart candidates, not automatically as confirmed physical associations. Rank candidates primarily by angular separation and avoid claiming a secure identification based only on positional proximity.

Return a summary table containing:

| Detected source | Image coordinates | RA | Dec | Candidate identifier | Object type | Angular separation | Search radius | Candidate status |

Use the following candidate-status values:
- candidate found;
- multiple candidates;
- no candidate within radius;
- coordinates unavailable.

Also report:
- the source-detection method used;
- the total number of detected catalogue entries;
- the adopted matching radius and how it was chosen;
- the number of sources with one or more positional candidates;
- the number of sources without candidates;
- links to the detected-source catalogue and full cross-match JSON/CSV artefacts.
\end{texthighlight}

\subsection{Catalogue comparison}
\label{app:catalog-comparison-prompt}

The following prompt was used for the independent source-extraction and catalogue-comparison demonstration:

\begin{texthighlight}
Detect radio sources in the attached FITS image independently using two different conventional radio source extraction algorithms. Configure both source finders to produce catalogues of source components containing both image and celestial coordinates and available source properties, including peak flux density, integrated flux density, and source size.

Cross-match the resulting source catalogues using sky coordinates with a fixed matching radius of 4 arcsec. Use this radius consistently for all associations and enforce one-to-one matching, selecting the closest valid counterpart when multiple candidates are available.

Identify:
- sources detected by both algorithms;
- sources detected only by the first algorithm;
- sources detected only by the second algorithm.

For every matched source pair, report the source identifiers from both catalogues, image coordinates when available, celestial coordinates, angular separation, and the source properties available from both finders.

Generate a merged source catalogue containing one entry for every unique catalogue source or matched source pair. Preserve:
- the original identifiers from both catalogues;
- the provenance of every detection;
- whether the entry is common to both finders, unique to the first algorithm, or unique to the second algorithm;
- the original measurements from each contributing catalogue.

Compute and report summary statistics including:
- total number of detections from each source finder;
- number of common detections;
- number of detections unique to each source finder;
- matching fraction for each catalogue;
- mean, median, minimum, and maximum angular separation of matched pairs;
- precision, recall, and F1 score when the first algorithm is treated as the reference catalogue.

Treat catalogue matches as positional associations between finder outputs, not necessarily as confirmed one-to-one astrophysical source identifications. Note that differences may arise from source splitting, blending, centroid definition, and fitting strategy.

Return a summary table and save the complete comparison report and merged catalogue in JSON format.
\end{texthighlight}

\section{Literature-analysis benchmark}
\label{app:literature-benchmark}

\subsection{Representative benchmark questions}
\label{app:literature-questions}
The complete benchmark comprises questions spanning the five thematic areas described in Section~\ref{subsec:literature-analysis-tasks}. Representative examples are shown in the Table below, while the complete question set is provided in the project software repository.

\begin{table}[!ht]
\centering
\caption{Sample radio literature questions extracted from the benchmark dataset.}
\label{tab:literature_sample_questions}
\footnotesize%
\begin{tabular}{p{2.7cm}|p{1.4cm}|p{12.8cm}}
\hline
\textbf{Thematic Area} & \textbf{ID} & \textbf{Question}\\
\hline
Basic knowledge & \texttt{BASE-001} & \textit{To what extent does incomplete (uv) coverage in radio interferometry introduce systematic biases into the morphological and spectral reconstruction of diffuse emission relative to compact components in the same field?}\\%
\hline%
Methods and techniques & \texttt{METH-009} & \textit{How do residual calibration errors propagate into spectral indices, rotation measures, fractional-polarization measurements, and subsequently inferred physical parameters?} \\%
\hline%
Galactic astronomy & \texttt{GAL-001} & \textit{How many Galactic supernova remnants remain absent from current catalogs, and which combinations of radio, infrared, gamma-ray, and X-ray observations are most effective for recovering them?} \\%
\hline%
Extragalactic astronomy & \texttt{EGAL-016} & \textit{To what extent can multiwavelength spectral energy distributions of blazars constrain particle acceleration and emission mechanisms in relativistic jets, and what are the main model-dependent uncertainties?} \\%
\hline%
Open questions & \texttt{OQ-002} & \textit{Which multiwavelength observable would provide the most decisive evidence that FRBs are dominated by young magnetars, multiple progenitor channels, or a still-unknown source population?} \\%
\hline%
\hline%
\end{tabular}
\end{table}

\subsection{Literature-analysis prompt}
\label{app:literature-analysis-prompt}
The following prompt was used for the literature-analysis benchmark. The placeholder \texttt{{QUESTION}} was replaced by each benchmark question, while \texttt{{DOMAIN\_DESCRIPTION}} specified the corresponding scientific domain.

\begin{texthighlight}
## Role
You are a scientific literature analyst specializing in {DOMAIN_DESCRIPTION}.
Your task is to answer technical questions by synthesizing published scientific evidence. Do not produce a generic textbook response when the literature is more nuanced or conditional.

## Question
{QUESTION}

## Scope
- Intended audience: researchers or advanced graduate students
- Literature cutoff date: More recent than year 2000
- Target length: 250--300 words
- Hard maximum: 350 words, excluding references

## Instructions
1. Answer the question directly in the opening paragraph. Do not begin with a broad introduction unless it is necessary to interpret the question.

2. Identify the central task posed by the question and answer it explicitly. Depending on the question, this may require explaining a mechanism, comparing alternatives, assessing the magnitude of an effect, summarizing evidence, identifying dependencies, or describing limitations. Include only the dimensions that are relevant to the question.

3. Explain the relevant physical, observational, or methodological mechanisms needed to understand the answer.

4. Distinguish clearly between:
- conclusions directly supported by observations, experiments, simulations, or formal analysis;
- interpretations or reasonable inferences;
- unresolved, disputed, or insufficiently studied aspects.

5. Report quantitative results when they are supported by the literature. Always state the conditions under which a numerical result applies. Do not generalize a result beyond the observational, instrumental, or methodological setting in which it was obtained.

6. Identify the assumptions, dependencies, and limitations that materially affect the answer.

7. Discuss mitigation strategies, methodological recommendations, or practical implications only when relevant to the question.

8. Prefer a synthesis of the evidence over a paper-by-paper summary.

## Evidence and citation rules
1. Base substantive factual claims only on the sources supplied through the MAASAI literature-retrieval context.

2. Cite sources inline using the supplied source identifiers. Place each citation immediately after the claim or group of claims it supports.

3. Do not cite a source unless it directly supports the associated statement.

4. Do not invent, infer, or reconstruct authors, titles, publication years, journal names, page numbers, DOIs, or other bibliographic information.

5. Every numerical or configuration-specific claim must have a citation.

6. When sources disagree, describe the disagreement and, where possible, explain whether it arises from different data, assumptions, definitions, or methods.

7. When the available sources support only part of the answer, state exactly which part is unsupported or uncertain.

8. When the supplied evidence is insufficient to answer the question, state: "The supplied literature is insufficient to determine [...]". Do not fill the gap using unsupported speculation.

9. Do not treat the absence of a result in the supplied sources as evidence that the effect or relationship does not exist.

## Required output

### Answer
Provide a coherent and concise synthesis in paragraphs. Include, where relevant:
- the main conclusion;
- the supporting evidence;
- the relevant mechanism;
- important dependencies or qualifications;
- uncertainty, disagreement, or evidence limitations;
- practical implications or mitigation strategies.

Do not force these elements into the answer when they are not relevant to the question.

### References
List only the sources cited in the answer, in order of first appearance.
Use only bibliographic metadata supplied by the retrieval context.
Do not include uncited sources.

### Final validation
Before returning the answer, silently verify that:
- the response directly addresses the actual question;
- all material parts of the question have been considered;
- every citation maps to a supplied source;
- each cited source supports the associated claim;
- every source in the reference list is cited in the answer;
- no bibliographic metadata has been invented;
- numerical claims are cited and appropriately qualified;
- uncertainty is proportional to the available evidence;
- the response remains within the requested length.

Return only the Answer and References sections.
\end{texthighlight}

\subsection{Literature-answer evaluation}
\label{app:literature-evaluation-prompt}
The \maasai{} literature-analysis responses were evaluated using the same five criteria for both the automated LLM judge and the human assessment by domain experts. Evaluators were provided with the benchmark question, the \maasai{} response, and the literature sources retrieved by the system, including the corresponding retrieved passages when available. Each criterion was scored independently on an integer scale from 0 to 5.

\begin{texthighlight}
You are evaluating a scientific literature-analysis response produced by an AI system.

You are given a scientific question, the response produced by the system, and the literature sources retrieved by the system, including retrieved passages when available.

Evaluate the response using exactly the five criteria below. Apply the rubric consistently and score each criterion independently.

## General evaluation principles

* Assess the response at the level expected of a researcher or advanced graduate student.
* Assess scientific correctness using your scientific expertise.
* Assess literature support only against the supplied sources and retrieved evidence.
* Do not assume that a claim is supported merely because it is scientifically plausible.
* Do not penalize the response under literature support merely because the retrieved literature is incomplete when the response explicitly and appropriately acknowledges that limitation.
* Do penalize unsupported claims that are presented as established conclusions.
* Do not reward verbosity. A concise answer can receive the maximum score if it adequately addresses the question.
* Do not penalize stylistic differences that do not affect scientific quality.
* Evaluate only the response provided; do not rewrite or improve it.
* Use integer scores only.

## Common scoring scale

0 = unacceptable

1 = major problems

2 = substantial problems

3 = adequate

4 = good, with only minor problems

5 = excellent

Use 1, 2, and 4 for intermediate cases between the criterion-specific anchors below.

### 1. Scientific correctness

Question: Is the scientific content accurate and are the conclusions justified?

0: Fundamentally incorrect or misleading.

3: Generally correct, but contains some errors, questionable statements, or unjustified conclusions.

5: Scientifically accurate, with appropriately justified conclusions.

### 2. Relevance and completeness

Question: Does the response directly answer the question and cover its important aspects?

0: Fails to address the central question.

3: Addresses the main question but omits or inadequately treats some relevant aspects.

5: Directly and comprehensively addresses the important aspects of the question.

### 3. Literature support

Question: Are substantive claims supported by the supplied literature, and do the citations support the statements with which they are associated?

0: Important claims are largely unsupported, or citations substantially misrepresent the supplied evidence.

3: Main claims are supported, but some claims have missing, weak, incomplete, or imperfect citation support.

5: Important claims are well supported and citations accurately trace them to the supplied evidence.

### 4. Synthesis quality

Question: Does the response integrate the available evidence into a coherent scientific answer rather than merely listing information or individual studies?

0: The response is incoherent, poorly reasoned, or lacks meaningful synthesis.

3: Provides a reasonable synthesis but with limited integration, comparison, or interpretation of the evidence.

5: Provides a clear and concise scientific synthesis that effectively integrates the relevant evidence.

### 5. Uncertainty and limitations

Question: Does the response appropriately represent uncertainty, disagreement, assumptions, and limitations when these are relevant to the question and available evidence?

0: Seriously overstates conclusions or ignores important uncertainties or limitations.

3: Generally represents uncertainty appropriately but misses some relevant qualifications or limitations.

5: Represents uncertainty, disagreement, and limitations accurately and proportionately.

N/A: Use only when uncertainty or limitations are genuinely not meaningful for the question or response.

## Comments

* For each criterion, provide at most one short sentence identifying the most important reason for the score.
* For a score of 5, the comment may be empty.
* Provide one brief overall comment focused on the main strength or weakness.
* Do not provide an alternative answer to the scientific question.
\end{texthighlight}

\subsection{Statistical analysis of human and automated evaluations}
\label{app:literature-evaluation-statistics}

Agreement between human reviewers and the automated LLM judge was assessed separately for each of the five evaluation criteria. Because the scores are ordinal integers ranging from 0 to 5, we considered both direct score differences and statistics accounting for chance agreement or ordinal association.

\paragraph{Inter-human reliability}
We used Krippendorff's ordinal alpha ($\alpha_{\mathrm{ordinal}}$) \citep{Krippendorff2013} to quantify the reproducibility of human ratings assigned to the same answers. This coefficient can accommodate different numbers of reviewers per answer and
missing ratings, and compares the observed disagreement $D_{o}$ among reviewers with the disagreement $D_{e}$ one would expect under chance-expected agreement, using distances appropriate for ordinal categories (see the original reference or auxiliary guides\footnote{\url{https://www.asc.upenn.edu/sites/default/files/2021-03/Computing\%20Krippendorff\%27s\%20Alpha-Reliability.pdf}}):

\begin{equation}
\alpha_{\mathrm{ordinal}} = 1 - \frac{D_o}{D_e}.
\label{eq:krippendorff-alpha}
\end{equation}

An alpha of 1 indicates perfect agreement between reviewers, whereas 0 indicates agreement no better than expected by chance; negative values indicate greater disagreement than expected by chance. Low alpha values therefore indicate limited reproducibility of the assigned scores, rather than necessarily large numerical differences between individual ratings.

Only answers with at least two valid human ratings contributed to the inter-human reliability analysis. Uncertainty estimates were obtained using 1,500 question-level bootstrap resamples,
retaining the complete set of human ratings associated with each sampled answer. The resulting 95\% percentile confidence intervals quantify the uncertainty of the estimated agreement coefficients.

\paragraph{Human--LLM agreement}
Agreement between individual human ratings and the automated LLM judge was assessed using the proportion of exact score matches, the proportion of scores differing by at most one point, and
quadratic weighted Cohen's kappa ($\kappa_w$) \citep{Cohen1968}. The latter is a chance-corrected agreement coefficient that accounts for the magnitude of differences between ordinal ratings.

Let $O_{ij}$ denote the observed relative frequency of pairs in which a human reviewer assigns score $i$ and the LLM judge assigns score $j$, and let $E_{ij}$ denote the corresponding relative frequency expected under independence of their marginal score distributions. Quadratic weighted kappa is
defined as\footnote{Equivalently, the coefficient can be expressed as $\kappa_w=(p_o-p_e)/(1-p_e)$, where $p_o$ and $p_e$ denote the observed and chance-expected weighted agreement.}:

\begin{equation}
\kappa_w =
1 -
\frac{\displaystyle\sum_{i=0}^{5}\sum_{j=0}^{5}
w_{ij} O_{ij}}
{\displaystyle\sum_{i=0}^{5}\sum_{j=0}^{5}
w_{ij} E_{ij}},
\label{eq:weighted-kappa}
\end{equation}

where the disagreement weights are: $w_{ij} = \frac{(i-j)^2}{25}$. Here, $w_{ij}=0$ for identical scores and increases quadratically with their difference, reaching unity for the maximum possible disagreement on the 0--5 scale. As for Krippendorff's alpha, a coefficient of 1 indicates perfect agreement, 0 corresponds to chance-expected agreement, and negative values indicate greater disagreement than expected by chance.

For our analysis, each valid individual human rating was paired with the LLM score assigned to the same answer. Consequently, answers evaluated by multiple astronomers contributed multiple pairs. The resulting pooled coefficient was interpreted descriptively, since the pairs are not statistically independent and the human reviewers vary across answers.

\paragraph{Rank association}
We additionally used Spearman's ($\rho$) \citep{Spearman1904} and Kendall's ($\tau_b$) \citep{Kendall1945} rank correlation coefficients to assess whether higher LLM scores tended to correspond to higher human scores across answers. For each answer, the LLM score was compared with
the median of the available valid human ratings, so that each answer contributed one observation. 

Unlike agreement coefficients, rank correlations assess relative ordering rather than equality of scores. For example, consistently assigning scores one point below the human ratings may preserve a perfect rank ordering despite imperfect absolute agreement. Correlations near zero indicate
little consistent monotonic association between the two evaluations, whereas values approaching 1 or -1 indicate strong positive or negative associations, respectively.

\paragraph{Scoring differences}
Potential directional scoring differences were assessed by subtracting the mean human rating from the corresponding LLM score for each answer. The overall mean difference was then calculated by averaging these answer-level differences, so that each answer contributed once, regardless of the number of human reviewers. Negative values indicate lower LLM scores, while positive values indicate higher LLM scores. Confidence intervals were obtained through question-level bootstrap resampling. Since the rating scale is ordinal, the numerical differences are interpreted as descriptive score differences rather than measurements on a strictly continuous scale.

\newpage%

\section{Supplementary figures}
\label{appendix:suppl-figs}
\setcounter{figure}{0}

\subsection{Image analysis tasks}
\vspace{-0.7cm}%

\begin{figure*}[!htbp]
\centering%
\subtable{\includegraphics[scale=0.37]{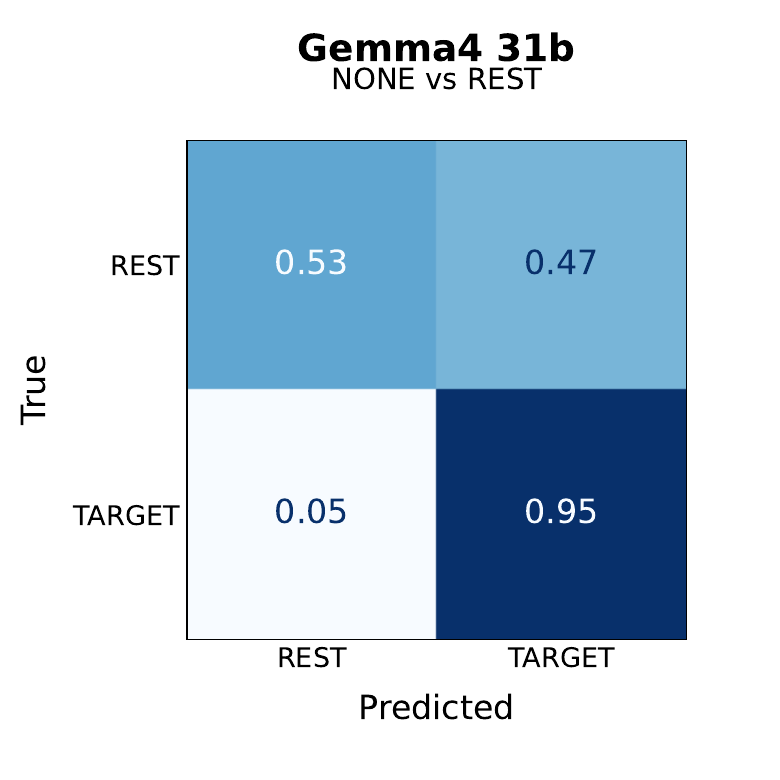}}
\hspace{-0.6cm}%
\subtable{\includegraphics[scale=0.37]{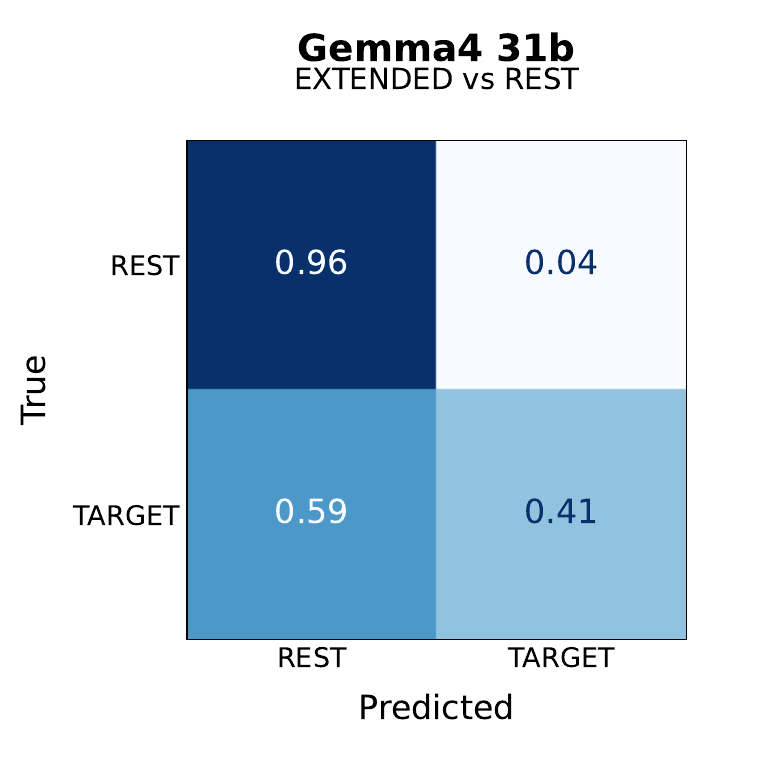}}
\hspace{-0.6cm}%
\subtable{\includegraphics[scale=0.37]{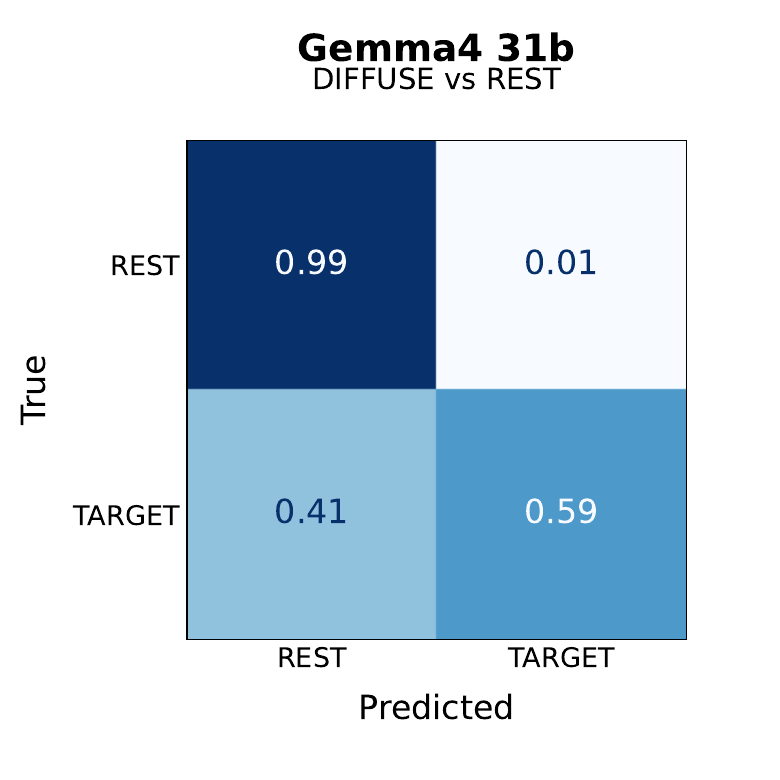}}
\hspace{-0.6cm}%
\subtable{\includegraphics[scale=0.37]{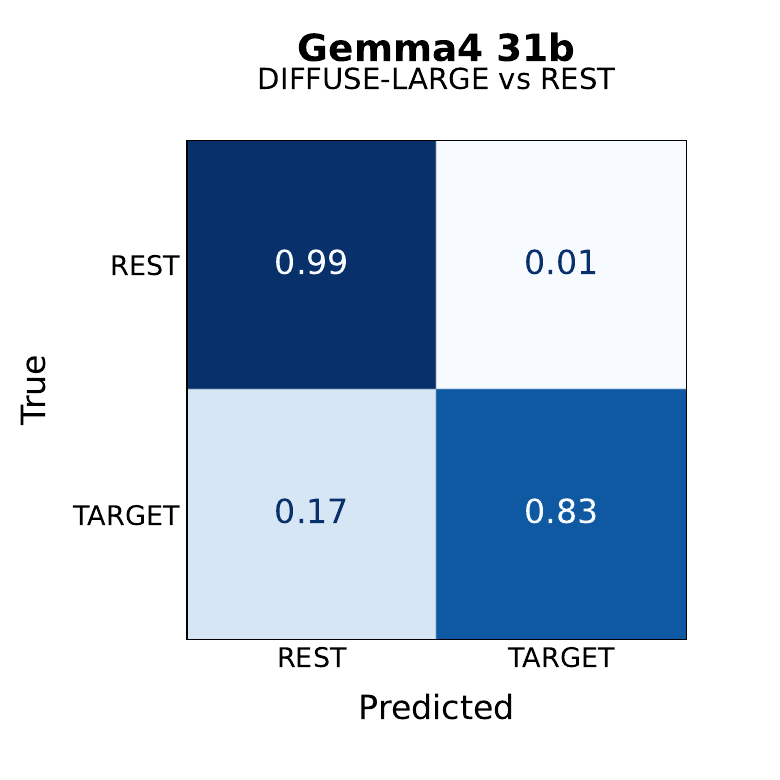}}\\%
\vspace{-0.7cm}%
\subtable{\includegraphics[scale=0.37]{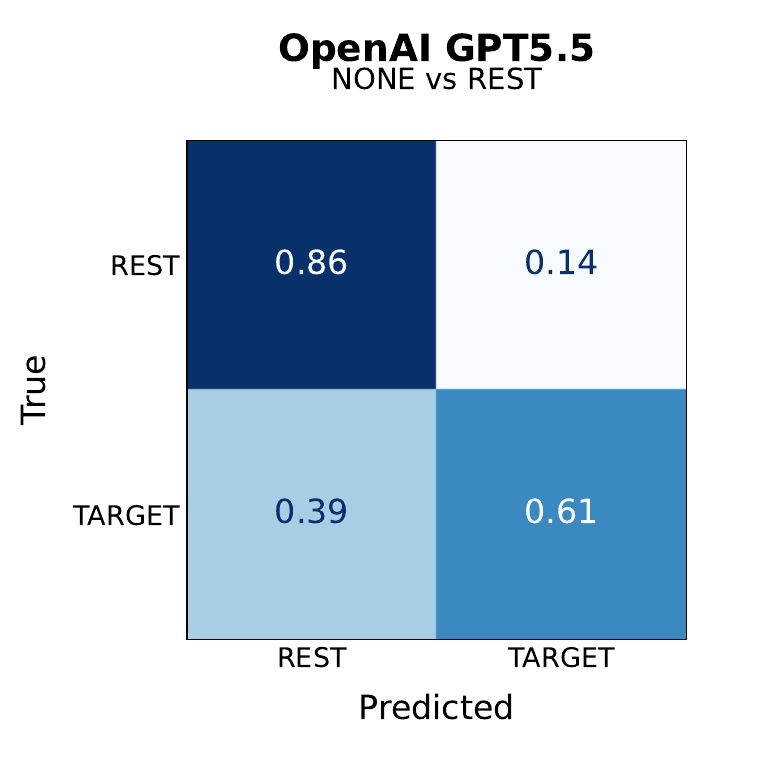}}
\hspace{-0.6cm}%
\subtable{\includegraphics[scale=0.37]{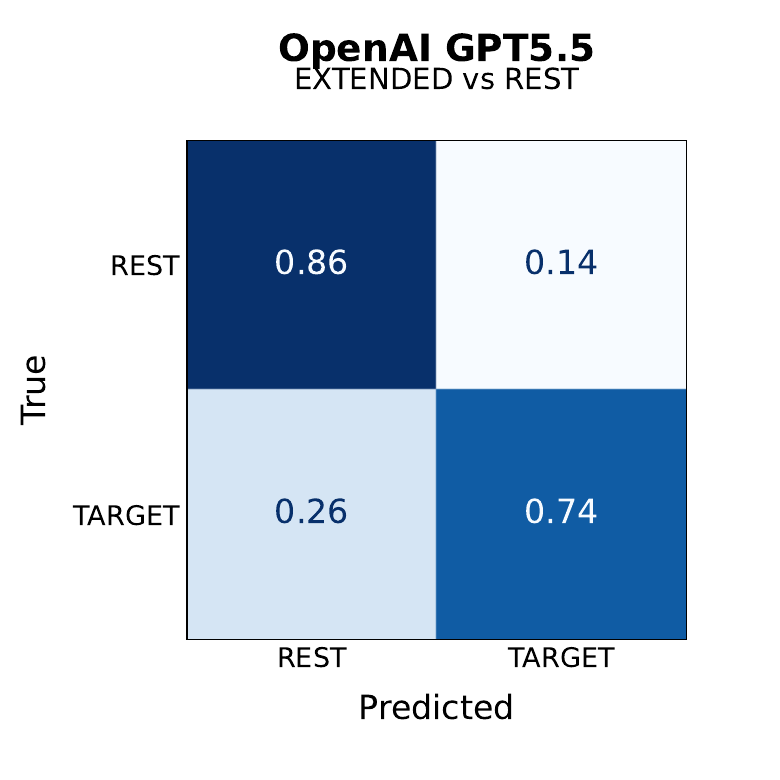}}
\hspace{-0.6cm}%
\subtable{\includegraphics[scale=0.37]{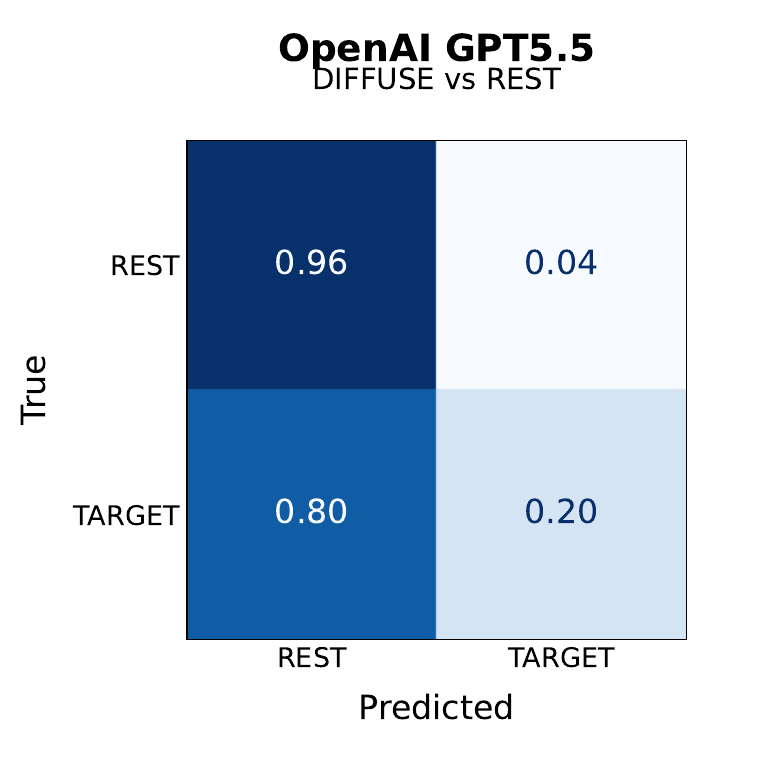}}
\hspace{-0.6cm}%
\subtable{\includegraphics[scale=0.37]{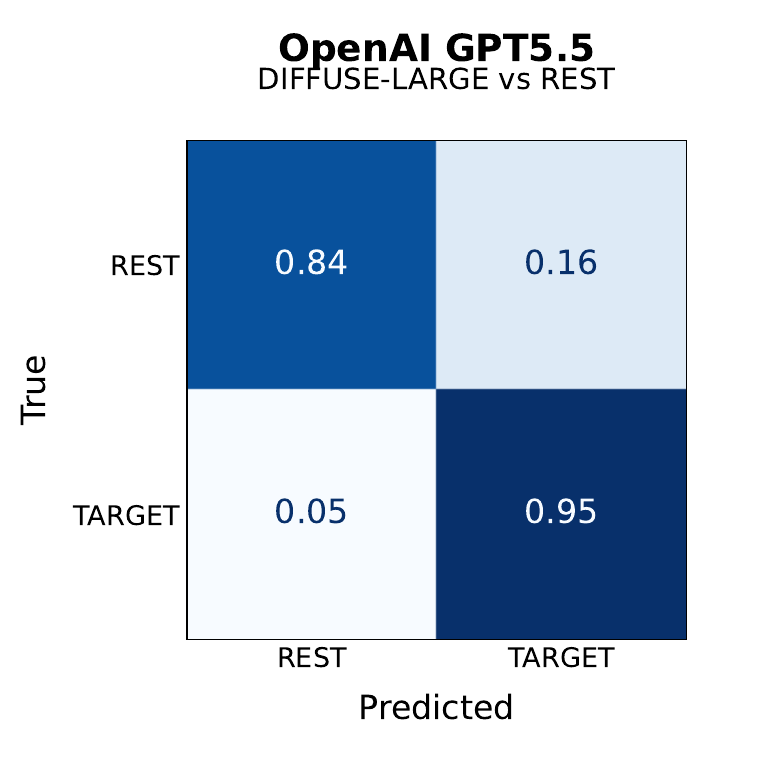}}\\%
\vspace{-0.8cm}%
\caption{Confusion matrices (one-class-vs-rest) obtained on the B1 radio benchmark (extended/diffuse source detection). Results obtained with the \maasai{} system, configured with a Gemma4:31b model, are shown in the top row, while those obtained with an OpenAI GPT baseline are shown in the bottom row.}%
\label{fig:cm_B1}
\end{figure*}


\begin{figure*}[!htbp]
\centering%
\subtable{\includegraphics[scale=0.5]{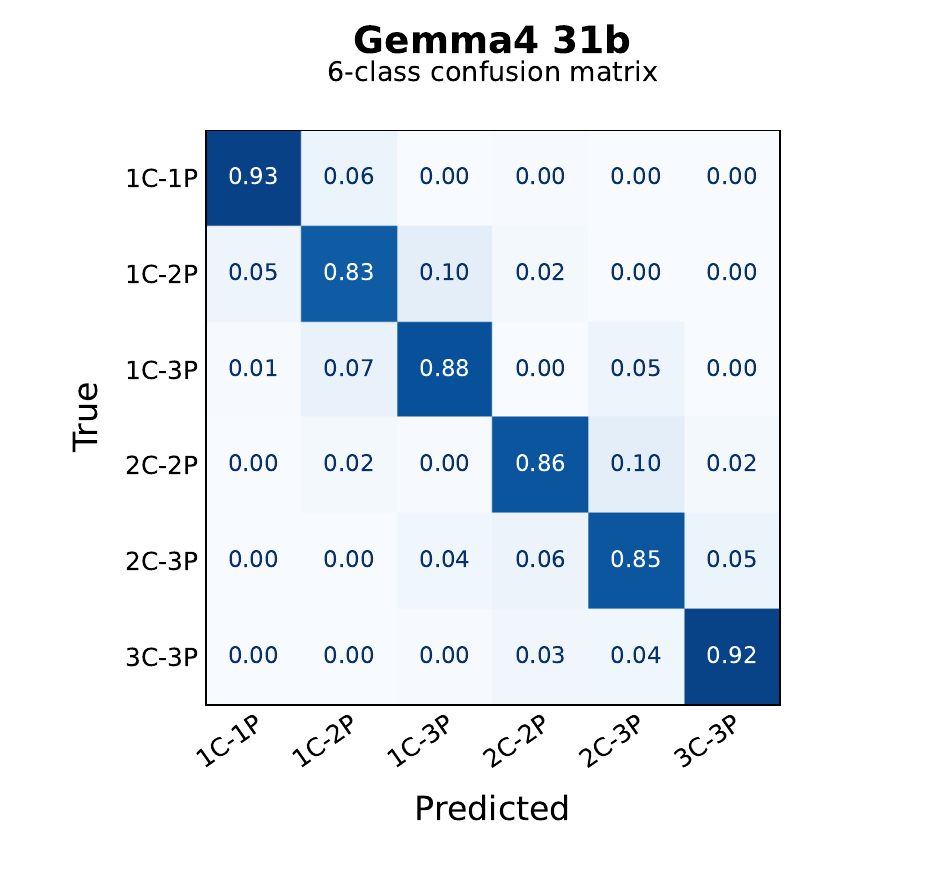}}
\hspace{-0.2cm}%
\subtable{\includegraphics[scale=0.5]{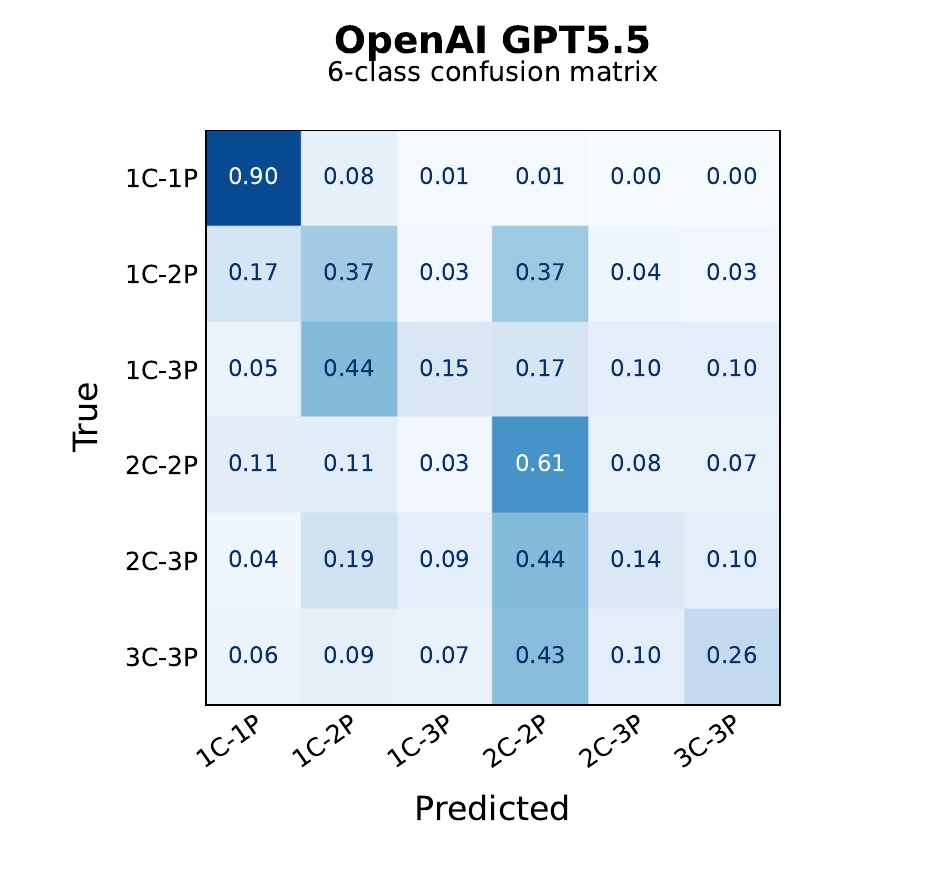}}%
\vspace{-0.7cm}%
\caption{Confusion matrices obtained on the B2 radio benchmark (source morphology classification). Results obtained with the \maasai{} system, configured with a Gemma4:31b model, are shown in the top row, while those obtained with an OpenAI GPT baseline are shown in the bottom row.}%
\label{fig:cm_B2}
\end{figure*}

\clearpage%

\subsection{Literature analysis tasks}

\begin{figure*}[!htbp]
\centering
\includegraphics[width=0.8\textwidth]{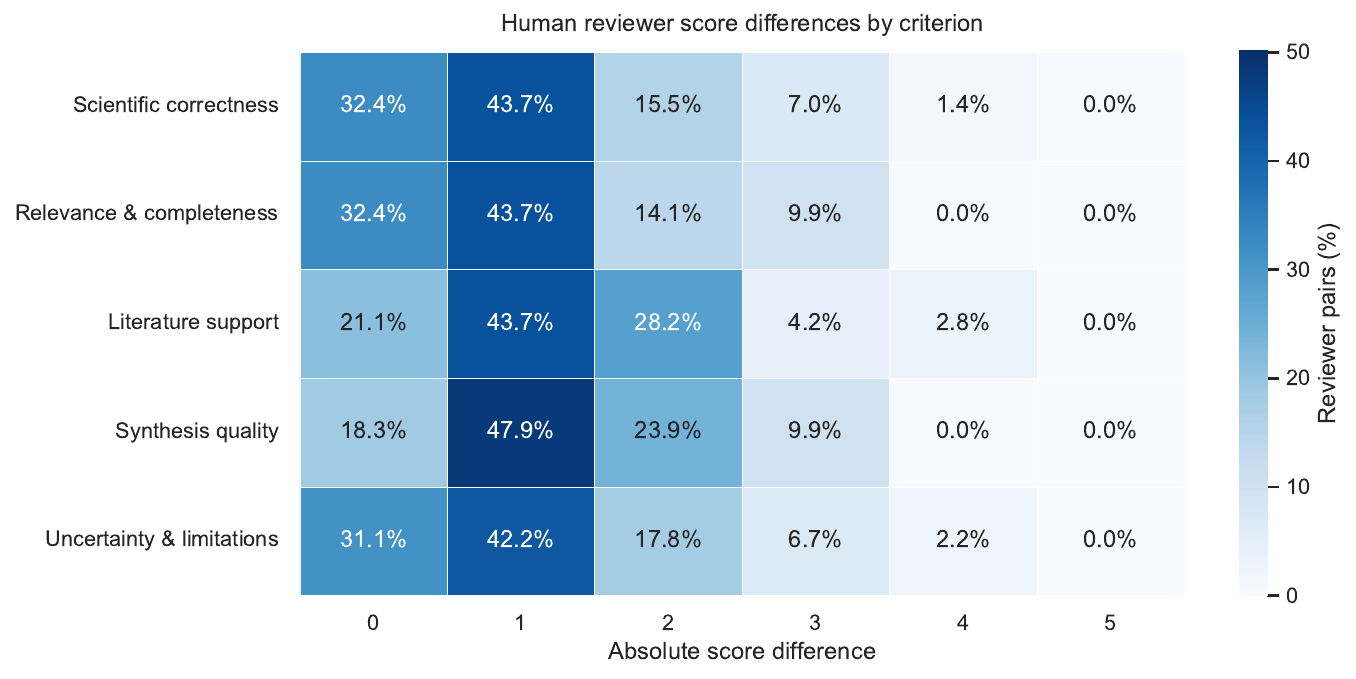}
\vspace{-0.4cm}%
\caption{Distribution of absolute score differences between human reviewer pairs across the five evaluation criteria. Each cell reports the percentage of distinct reviewer pairs with the corresponding absolute score difference, from 0 (exact agreement) to 5 points. The analysis includes 71 reviewer pairs across 28 answers for the first four criteria and 45 pairs across 20 answers for uncertainty and limitations. Percentages in the first column indicate exact agreement, while the sum of the first two columns corresponds to reviewer pairs differing by at most one score point. Pairwise comparisons are descriptive because multiple pairs may originate from the same answer and are not statistically independent.}
\label{fig:literature-human-pair-differences}
\end{figure*}

\vspace{-0.5cm}%

\begin{figure*}[!htbp]
\centering
\includegraphics[width=0.8\textwidth]{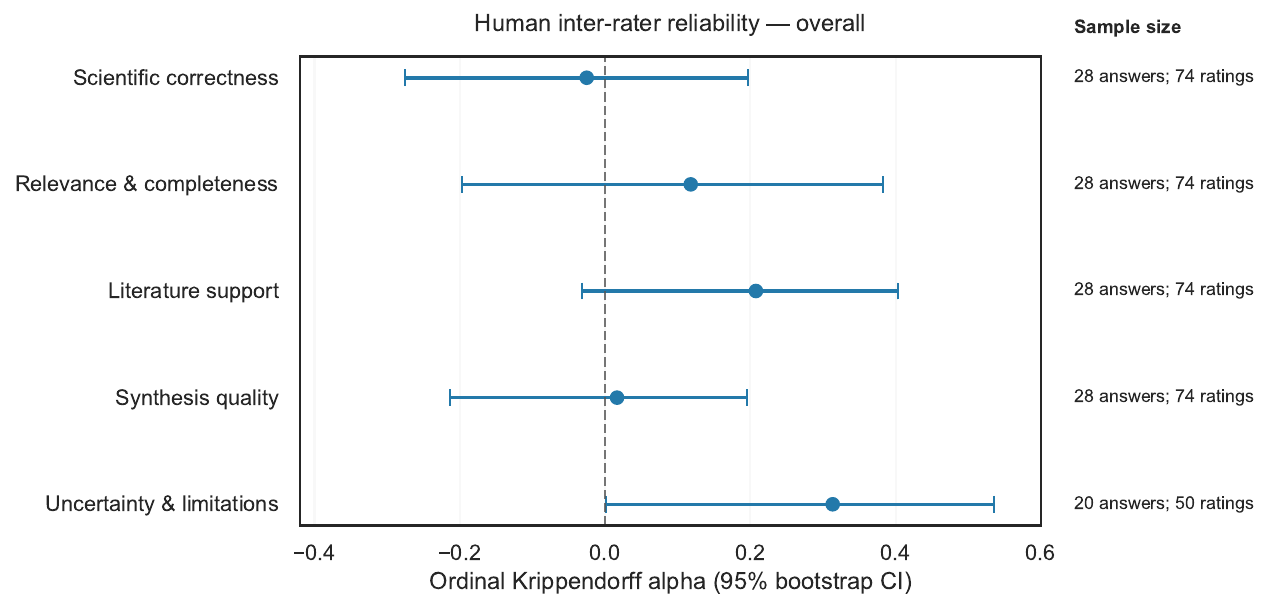}
\vspace{-0.2cm}%
\caption{Inter-human reliability across the five scientific evaluation criteria, measured using Krippendorff's ordinal alpha ($\alpha_{\mathrm{ordinal}}$). Points represent the estimated coefficients, with horizontal error bars indicating 95\% confidence intervals from 1,500 question-level bootstrap resamples. The dashed vertical line marks $\alpha_{\mathrm{ordinal}}=0$, indicating chance-level agreement. The number of contributing answers and valid human ratings is reported for each criterion.}
\label{fig:literature-human-alpha}
\end{figure*}

\begin{figure*}[!htbp]
\centering
\includegraphics[width=0.8\textwidth]{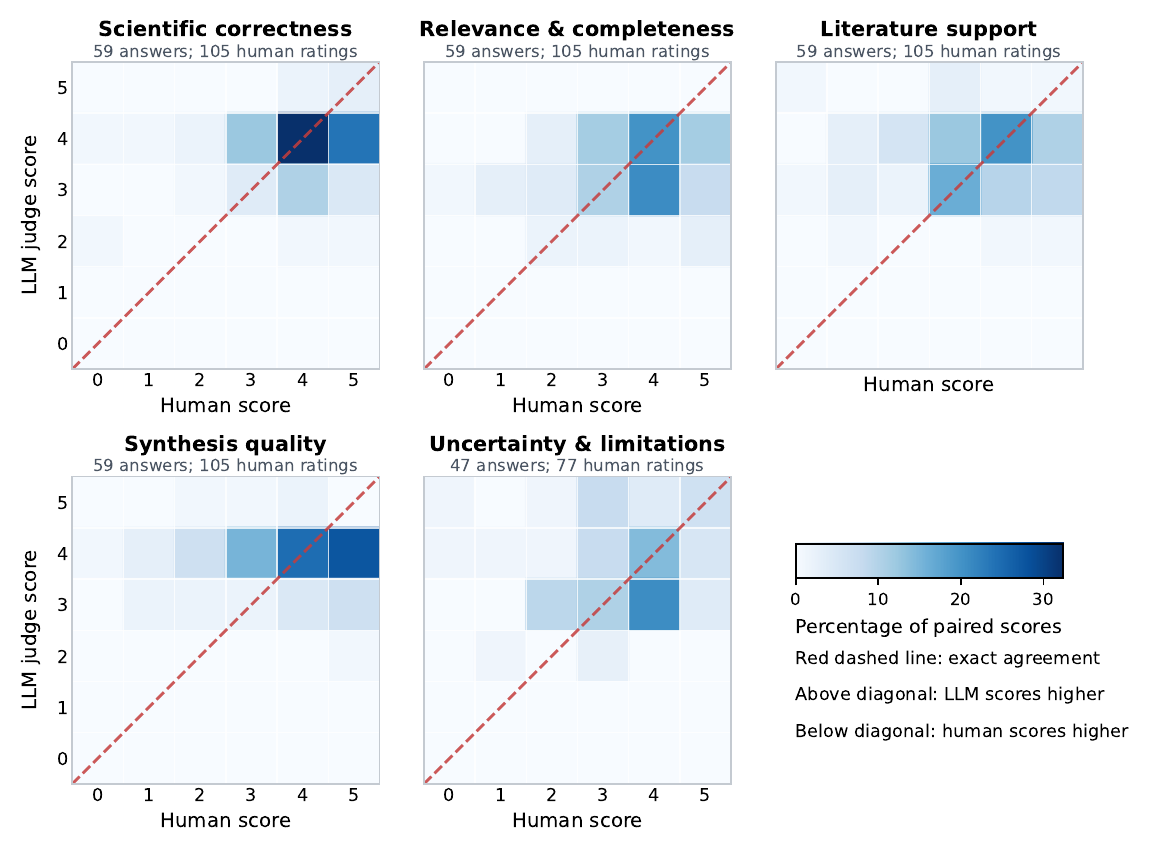}
\vspace{-0.4cm}%
\caption{Comparison of automated LLM judge scores and individual human ratings across the five literature-evaluation criteria. Each panel shows the distribution of paired scores on the 0--5
scale, with human ratings on the horizontal axis and LLM ratings on the vertical axis. Cell colours indicate the percentage of matched human--LLM rating pairs assigned to each score combination. The dashed diagonal represents exact agreement, while cells above and below it correspond to higher and lower LLM scores, respectively. The number of contributing answers and human ratings is indicated in each panel. Multiple pairs may originate from the same answer and are therefore not statistically independent}
\label{fig:literature-human-llm-score-map}
\end{figure*}

\begin{figure*}[!htbp]
\centering
\includegraphics[width=0.65\textwidth]{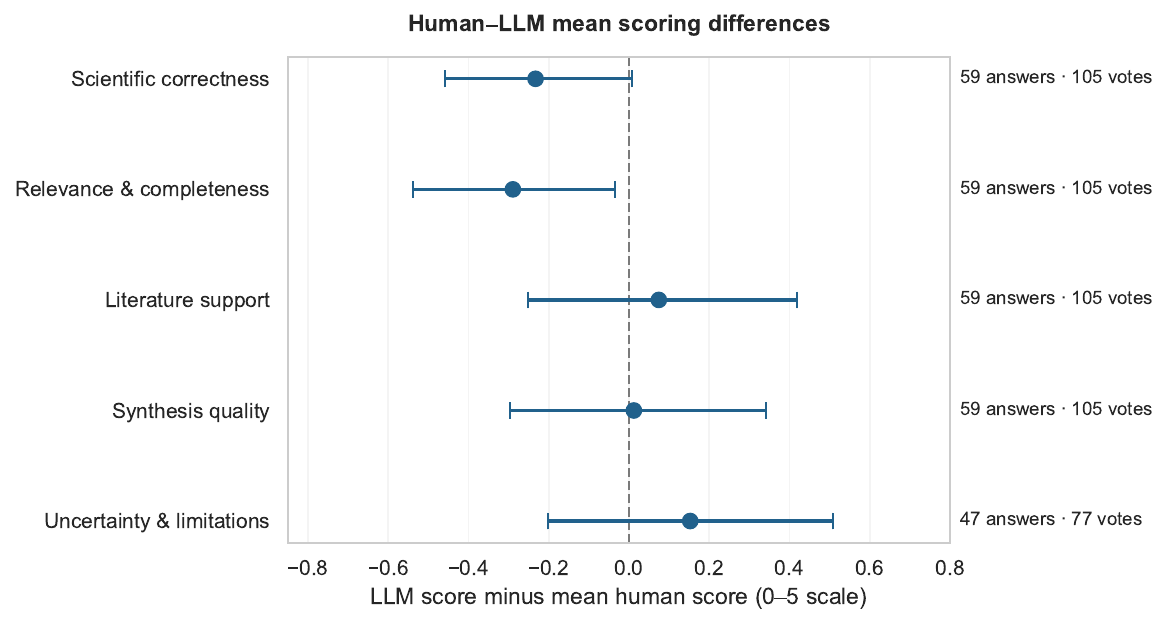}
\vspace{-0.2cm}%
\caption{Mean scoring differences between the automated LLM judge and human reviewers across the five evaluation criteria. For each answer, the difference is calculated as the LLM score minus the mean of its available human ratings, giving equal weight to each answer. Points represent the mean differences, with horizontal error bars indicating 95\% confidence intervals from 1,500 question-level bootstrap resamples. Negative and positive values indicate lower and higher LLM scores, respectively. The dashed vertical line marks zero mean difference.}
\label{fig:literature-human-llm-bias}
\end{figure*}

\begin{figure*}[!htbp]
\centering
\includegraphics[width=0.9\textwidth]{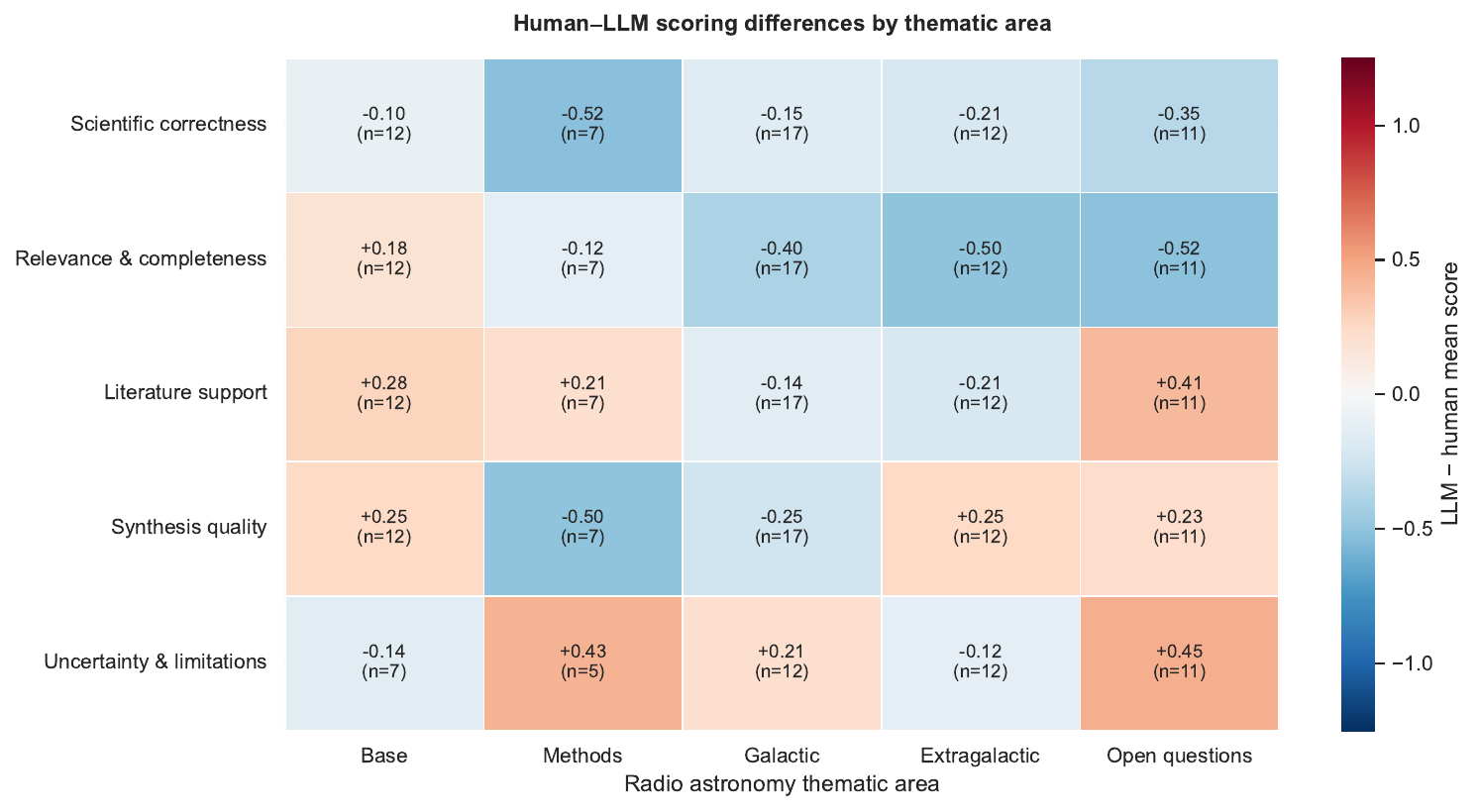}
\vspace{-0.2cm}%
\caption{Mean human--LLM scoring differences across the five radio-astronomy thematic areas and evaluation criteria. Each cell reports the mean difference between the LLM score and the mean human rating, assigning equal weight to each answer. The number of contributing answers is indicated in parentheses. Negative values indicate lower LLM scores, while positive values indicate higher scores. These domain-specific comparisons are exploratory and are not adjusted for multiple testing.}
\label{fig:literature-human-llm-bias-thematic}
\end{figure*}

\clearpage%

\section{Supplementary tables}
\label{appendix:suppl-tables}
\setcounter{table}{0}

{ 
\footnotesize%
\begin{longtable}{p{5cm}p{2.2cm}p{1.2cm}p{7.5cm}}
\caption{Current \maasai{} scientific tool inventory. Native tools are implemented directly within the \texttt{LangGraph} workflow, whereas external applications are exposed through the \texttt{caesar-rest} service infrastructure. The table includes applications available through the external tool registry, including capabilities not explicitly evaluated in the workflows presented in this work.}%
\label{tab:maasai_tools}\\

\hline%
{\small\textbf{Tool}} & {\small\textbf{Category}} & {\small\textbf{Backend}} & {\small\textbf{Purpose}} \\
\hline
\endfirsthead

\hline
{\small\textbf{Tool}} & {\small\textbf{Category}} & {\small\textbf{Backend}} & {\small\textbf{Purpose}}\\%
\hline
\endhead

\hline
\multicolumn{4}{r}{Continued on next page}\\
\endfoot

\hline
\endlastfoot

\multicolumn{4}{l}{\textbf{Image I/O and image analysis}}\\
\hline

{\scriptsize \texttt{fits\_image\_info\_tool}}
& {\scriptsize Image I/O}
& {\scriptsize Native}
& {\scriptsize Extract FITS metadata, WCS information, beam parameters, observing frequency and image statistics.} \\

{\scriptsize\texttt{find\_sources\_simple}}
& {\scriptsize Image analysis}
& {\scriptsize Native}
& {\scriptsize Fast threshold-based source detection producing lightweight source catalogs.} \\

{\scriptsize\texttt{measure\_sources\_from\_ds9\_regions}}
& {\scriptsize Image analysis}
& {\scriptsize Native}
& {\scriptsize Measure photometric and morphological properties inside DS9 source regions.} \\

\hline
\multicolumn{4}{l}{\textbf{Image cutout generation}}\\
\hline

{\scriptsize\texttt{image\_cutout\_tool}}
& {\scriptsize Cutouts}
& {\scriptsize Native}
& {\scriptsize Extract FITS/PNG cutouts around user-specified pixel or sky coordinates.} \\

{\scriptsize\texttt{source\_cutout\_tool}}
& {\scriptsize Cutouts}
& {\scriptsize Native}
& {\scriptsize Generate cutouts for all sources contained in a source catalog.} \\

{\scriptsize\texttt{region\_cutout\_tool}}
& {\scriptsize Cutouts}
& {\scriptsize Native}
& {\scriptsize Generate cutouts from DS9 regions with configurable margins.} \\

\hline
\multicolumn{4}{l}{\textbf{Source catalog analysis}}\\
\hline

{\scriptsize\texttt{catalog\_crossmatch\_tool}}
& {\scriptsize Catalog analysis}
& {\scriptsize Native}
& {\scriptsize Cross-match or merge two or more source catalogs using positional or bounding-box matching.} \\

{\scriptsize\texttt{catalog\_compare\_tool}}
& {\scriptsize Catalog analysis}
& {\scriptsize Native}
& {\scriptsize Compare catalogs and compute completeness, purity, positional offsets and flux statistics.} \\

{\scriptsize\texttt{catalog\_filter\_tool}}
& {\scriptsize Catalog analysis}
& {\scriptsize Native}
& {\scriptsize Select catalog entries according to user-defined filtering criteria.} \\

{\scriptsize\texttt{catalog\_statistics\_tool}}
& {\scriptsize Catalog analysis}
& {\scriptsize Native}
& {\scriptsize Compute descriptive statistics of source catalogs.} \\

{\scriptsize\texttt{catalog\_visualization\_tool}}
& {\scriptsize Visualisation}
& {\scriptsize Native}
& {\scriptsize Produce overlay images and statistical plots from one or more source catalogs.}\\

{\scriptsize\texttt{source\_catalog\_enhancement\_tool}}
& {\scriptsize Catalog enhancement}
& {\scriptsize Native}
& {\scriptsize Enhance traditional source catalogs using ML detector outputs by tagging, grouping and quality flagging detections.} \\

{\scriptsize\texttt{archive\_crossmatch\_tool}}
& {\scriptsize Catalog analysis}
& {\scriptsize Native}
& {\scriptsize Cross-match source catalogs against astronomical archive objects.} \\

\hline
\multicolumn{4}{l}{\textbf{Astronomical archive access}}\\
\hline

{\scriptsize Archive query tools}
& {\scriptsize Archive}
& {\scriptsize Native}
& {\scriptsize Query astronomical databases (e.g. SIMBAD, Vizier) and retrieve source information.} \\

{\scriptsize Archive cutout tools}
& {\scriptsize Archive}
& {\scriptsize Native}
& {\scriptsize Retrieve survey cutouts from remote services (e.g. SkyView, LoTSS, MAGPIS, IRSA, Herschel).} \\

\hline
\multicolumn{4}{l}{\textbf{External \texttt{caesar-rest} applications}}\\
\hline

{\scriptsize\texttt{caesar}}
& {\scriptsize Source finding}
& {\scriptsize External}
& {\scriptsize Radio compact and extended source extraction and characterization.} \\

{\scriptsize\texttt{caesar-yolo}}
& {\scriptsize Source detection}
& {\scriptsize External}
& {\scriptsize Deep-learning object detection and radio-source morphology classification.} \\

{\scriptsize\texttt{aegean}}
& {\scriptsize Source finding}
& {\scriptsize External}
& {\scriptsize Radio compact source extraction and characterization.} \\

{\scriptsize\texttt{vit\_classifier}}
& {\scriptsize Classification}
& {\scriptsize External}
& {\scriptsize Whole-image radio morphology, anomaly and artefact classification.}\\

{\scriptsize\texttt{vit\_classifier\_catalog}}
& {\scriptsize Classification}
& {\scriptsize External}
& {\scriptsize Per-source morphology classification using catalog-driven image cutouts.}\\

{\scriptsize\texttt{sfforecaster}}
& {\scriptsize Forecasting/Classification}
& {\scriptsize External}
& {\scriptsize Solar flare 24h forecasting using active region magnetograms.}\\

{\scriptsize\texttt{umap}}
& {\scriptsize Dimensionality reduction}
& {\scriptsize External}
& {\scriptsize Perform nonlinear dimensionality reduction of numerical feature vectors for visualization and exploratory analysis.}\\

{\scriptsize\texttt{hdbscan}}
& {\scriptsize Clustering}
& {\scriptsize External}
& {\scriptsize Perform density-based clustering of numerical feature vectors, identifying clusters and noise or unassigned samples.}\\

{\scriptsize\texttt{outlier-finder}}
& {\scriptsize Anomaly detection}
& {\scriptsize External}
& {\scriptsize Identify anomalous observations in numerical feature space using an outlier-detection algorithm.}\\

{\scriptsize\texttt{similarity-search}}
& {\scriptsize Similarity search}
& {\scriptsize External}
& {\scriptsize Retrieves similar astronomical sources through
nearest-neighbor searches in learned feature spaces.}\\

{\scriptsize\texttt{fextractor}}
& {\scriptsize Multi-modal feature extraction}
& {\scriptsize External}
& {\scriptsize Extracts feature vectors from astronomical images and time series using pretrained foundation models,
deep-learning encoders, and statistical methods.}\\

{\scriptsize\texttt{oxkat}}
& {\scriptsize Radio data calibration \& imaging}
& {\scriptsize External}
& {\scriptsize MeerKAT Measurement Set inspection and metadata retrieval, with ongoing integration of calibration and imaging pipeline operations.}\\

\end{longtable}
}

\end{document}